\documentclass[aps,amsmath,twocolumn,amssymb,floatfixng,showpacs, superscriptaddress,footinbib, longbibliography]{revtex4-1}

\usepackage{graphicx}
\usepackage{dcolumn}
\usepackage{multirow}
\usepackage{booktabs}
\usepackage{bm,color}
\usepackage{braket}
\usepackage{amsmath,amssymb}
\usepackage[colorlinks,linkcolor=blue,hyperindex,CJKbookmarks]{hyperref}
\usepackage{epstopdf}

\newcommand{\bs}[1]{\textcolor{black}{#1}}
\newcommand{\bl}[1]{\textcolor{black}{#1}}

\begin{document}
\title{Engineering skyrmion from spin spiral in transition metal multilayers} 
\author{Banasree Sadhukhan}
\email{banasres@srmist.edu.in}
\affiliation{Department of Physics and Nanotechnology, SRM Institute of Science and Technology, Kattankulathur, 603203, Chennai, Tamil Nadu, India}
\affiliation{Tata Institute of Fundamental Research, Hyderabad, Telangana 500046,  India}
\affiliation{Department of Applied Physics, School of Engineering Sciences, KTH Royal Institute of Technology, AlbaNova University Center, SE-10691 Stockholm, Sweden}

\begin{abstract}
 
Skyrmions having topologically protected field configurations with particle-like properties play an important role in {\bl{various fields of science}}.  Our present study focus on the generation of skyrmion from spin spiral in the magnetic multilayers of 4d/Fe/Ir(111) with 4d = Y,  Zr,  Nb,  Mo,  Ru, Rh.  Here we investigate the impact of 4d transition metals on the isotropic Heisenberg exchanges and anti-symmetric Dzyaloshinskii-Moriya interactions originating from the broken inversion symmetry at the interface of 4d/Fe/Ir(111) multilayers.  We find a strong exchange frustration due to the hybridization of the Fe-3d layer with both 4d and Ir-5d layers which modifies due to band filling effects of the 4d transition metals.  We strengthen the analysis of exchange frustration by shedding light on the orbital decomposition of isotropic exchange interactions of Fe-3d orbitals.  Our spin dynamics and Monte Carlo simulations indicate that the magnetic ground state of 4d/Fe/Ir(111) transition multilayers is a spin spiral in the $ab$-plane with a period of 1 to 2.5 nm generated by magnetic moments of Fe atoms and propagating along the $a$-direction.  The spiral wavelengths  in Y/Fe/Ir(111)  are much larger compared to Rh/Fe/Ir(111).  In order to manipulate the skyrmion phase in 4d/Fe/Ir(111),  we investigate the magnetic ground state of 4d/Fe/Ir(111) transition multilayers with different external magnetic field.  An increasing external magnetic field of $\sim$ 12 T is responsible for deforming the spin spiral into a isolated skyrmion which flips into skyrmion lattice phase around $\sim$ 18 T in Rh/Fe/Ir(111).  Our study predict that the stability of magnetic skyrmion phase in Rh/Fe/Ir(111) against thermal fluctuations is upto temperature T $\leq 90$ K.  

\end{abstract}

\maketitle

\section{Introduction}

\par A magnetic skyrmion is a topologically protected spin structure that attracts giant attention due to its inherent magnetic stability against random thermal fluctuations \cite{Heinze2011-ke, Romming2013-uw,  Wiesendanger2016-zn}.   They are small in size,  typically ranging from a few nanometers to a few tens of nanometers and can be manipulated with low current densities \cite{Wiesendanger2016-zn,  Yu2012-yw}.  Small size,  stability,  low power requirements, and fast dynamics make skyrmions promising candidates for modern spintronics and high density data storage \cite{Fert2013-wu,  Nagaosa2013-fx,  Kiselev_2011, Romming2013-uw,  Jonietz2010-bo,  Schulz2012-ax,  Yu2012-yw,  Iwasaki2013-gi,   PhysRevB.87.214419,   Iwasaki2013-fg,  Zhang2015-mn,  shu2022realization, PhysRevB.98.134448,  Nagaosa2013-ns}.  Magntic skyrmions first {\bl {observed}} experimentally in bulk magnet MnSi with broken inversion symmetry \cite{Muhlbauer2009-hu}. Skyrmions can be stabilized experimentally in various systems such as bulk crystals \cite{Fert2013-wu,  Yu2010-ea,  PhysRevLett.107.127203,  PhysRevB.81.041203},  crystal thin films \cite{Heinze2011-ke,  Tonomura2012-mp,  Yu2011-ix},  ultrathin films \cite{Romming2013-uw,  Heinze2011-ke} , two-dimensional van der Waals magnets \cite{PhysRevB.102.241107,  Li2022-tx,  PhysRevB.107.054408}. The search continues to find materials/systems hosting skyrmions for its suitable applications.

\par  Magnetic multilayers are another class of systems in which chiral magnetic structures occur \cite{Wiesendanger2016-zn,  Bode2007-ku}.  Here,  thin {\bl{films}} of magnetic transition-metals are placed on heavy metal substrates (e.g., Pt, Ir, W) to provide strong spin-orbit interaction which induces noncollinear magnetic structures \cite{PhysRevB.97.134405,  PhysRevB.96.094408,  PhysRevB.98.060413,  PhysRevB.104.024420,   PhysRevB.107.174430,  PhysRevLett.120.207201}.  For example,  spin spiral (SS) phases have been reported in Mn-3d monolayer on W(110)/W(100) \cite{Bode2007-ku,  PhysRevB.91.064402,  PhysRevLett.101.027201}, whereas skyrmion lattice (SkL) phases have discovered in Fe-3d monolayer on the Ir(111) surface without external magnetic field using spin-polarized scanning tunnelling microscopy \cite{Heinze2011-ke}.  However,  a nonmagnetic overlayer deposited on 3d transition metal films such as Pd-Fe atomic bilayer on Ir(111) flips the magnetic ground state from SkL into SS \cite{Dupe2014-ej}. This opens the door of a new class of transition-metal multilayers where noncollinear spin texture from SS to skyrmion, and thereby the properties of skyrmions can be tailored \cite{PhysRevB.107.174430}.

\par A recipe to produce skyrmions in thin multilayers is the application of external magnetic fields that cause the spirals to evolve into skyrmions.  {\bs {Pd/Fe/Ir(111) is the most intensively studied multilayers system with a SS ground state and hosts isolated skyrmions with small diameters of about 4-6 nm at external magnetic field of $\approx$ 3-6 T \cite{Romming2013-uw,  Dupe2014-ej,  PhysRevB.90.094410,  sadhukhan2023spin,  PhysRevB.105.224413,  vonMalottki2017}. }} The existence of  SS or chiral magnetic skyrmions  are {\bl {attributed to}} the competition between the Heisenberg exchange interactions and Dzyaloshinskii-Moriya interaction (DMI),  which occurs due to spin-orbit coupling (SOC) \cite{PhysRev.120.91,  DZYALOSHINSKY1958241,  CREPIEUX1998341,  Bode2007-ku,  PhysRevB.110.174412}.  A critical value of DMI (D$_c$) is needed to stabilize the noncollinear spin structures which depends on the spin stiffness (A) and effective magnetocrystalline anisotropy energy (MAE) K given by $\mathrm{D_c} \propto \sqrt{\mathrm{AK}}$ \cite{BOGDANOV1994255,  Rosler2006-jf}.  The large DMI in Pd/Fe/Ir(111) occurs due to strong SOC originating from the 5d transition metal with the broken inversion symmetry at the interface.

\par Another approach is to consider 3d/4d interfaces instead of 3d/5d interface \cite{PhysRevB.96.094408,  Dupe2016-wp,  PhysRevLett.117.157205,  PhysRevB.90.094410,  Dupe2014-ej,  PhysRevB.90.020402,   PhysRevB.107.174430}.  The magnetic interactions at interfaces can be tuned by the hybridization between 3d/5d and 4d/5d-transition-metal multilayers \cite{PhysRevB.79.094411}. This tailors the noncollinear spin structures at transition-metal surfaces and interfaces.  Recently,  skyrmion phase has been reported in Pd-Fe/Rh(111) and Co/Ru(0001) instead of Ir(111) substrates \cite{PhysRevB.98.060413,  Herve2018-qo}.  In such systems,  both the DMI and MAE reduce due to a lower SOC.  Magnetic exchange frustration has been observed in both Rh-Fe and Pd-Fe bilayers on Ir(111), Rh/Co/Ir(111) \cite{PhysRevB.107.174430,  Meyer2019-ui,  PhysRevLett.120.207201}.  This leads to a SS  as magnetic ground state in both Rh/Fe/Ir(111) and Pd/Fe/Ir(111) \cite{PhysRevLett.120.207201,  Dupe2014-ej}.  Whereas,  Rh-Co/Ir(111) induces isolated skyrmions with diameters of  $\leq$ 10 nm at zero magnetic field \cite{Meyer2019-ui}.  Moreover,  the interfacial DMI stabilizes skyrmions in multilayers \cite{Dupe2014-ej}.

\par  Motivated by the above experimental and theoretical reports,  here,  we study the effect of different 4d transition metals on the magnetic exchange interactions and interfacial DMI in 4d/Fe/Ir(111) with 4d = Y, Zr, Nb,  Mo, Ru and Rh.  We demonstrate how the electronic hybridization between 4d transition metals with different numbers of valence electrons and Fe-3d affects the magnetic ground state of 4d/Fe/Ir(111) multilayers which is a SS.   We observe the magnetic exchange interactions are highly frustrated in 4d/Fe/Ir(111) transition metal multilayers and a significant amount of frustration leads to SkL phase from SS phase in Rh/Fe/Ir(111) with external magnetic field.  We explain our findings by the orbital decomposition of complex magnetic interactions and magnetic moments of Fe-3d due to 3d/4d hybridization.

\par The paper is structured as follows: In Sec.~\ref{sec_II},  we present the necessary theoretical and computational details.  In Sec.~\ref{sec_III},  we first discuss the calculated magnetic exchange interactions along with orbital analysis and interfacial DMI in 4d/Fe bilayers on Ir(111).  Then we determine the magnetic ground state of 4d/Fe/Ir(111) multilayrs which is a SS of wave length  $\lambda$  $\sim$ (1-2.5) nm from both spin dynamics and Monte Carlo simulations.   The period of spiral decreases if we move from Y to Rh in 4d/Fe/Ir(111). The magnetic ground state shows a phase transition from SS to SkL in Rh/Fe/Ir(111) with external magnetic field of  $\sim$ 18 T.   Moreover,  we also investigate the temperature stability of the magnetic skyrmion from Monte Carlo simulations which shows that skyrmion phase in Rh/Fe/Ir(111) is stable against thermal fluctuation upto T $\leq 90$ K.  Finally the results are summarized and ended by the  conclusion in Sec.~\ref{sec_IV}.

\section{Computational details}
\label{sec_II}

\par The density functional theoretical (DFT) calculations have been performed into two steps.  First,  we use fcc stacking of 4d/Fe bilayers on five layers of Ir(111) and add a vacuum of 20 Å on 4d/Fe/Ir(111) to make it as thin {\bl{ film}}.  {\bl{Here we consider fcc stacking of 4d/Fe/Ir(111) instead of hcp stacking because only fcc stacking of Pd/Fe/Ir(111) gives SKL phase with magnetic field \cite{vonMalottki2017}.}}  The in-plane lattice constant of the Ir(111) is 2.74 Å for modeling the geometry of 4d/Fe/Ir(111) multilayers.  Then we relaxed the structure using the projector augmented wave method (PAW) as implemented in VASP \cite{PhysRevB.59.1758,  PhysRevB.54.11169,  vasp}.  The plane-wave energy cutoff was set to 380 eV and the k-point mesh consists of 18$\times$18$\times$1.   Electronic structure are calculated using spin-polarized local spin density approximation (LSDA) implemented in VASP \cite{PhysRevB.59.1758,  PhysRevB.54.11169,  vasp}.   

\par To obtain the magnetic interactions of multilayer systems,  we have performed DFT total energy calculations with collinear configurations using LKAG formalism as implemented in RSPt code \cite{wills2010full, rsptweb}.  The radii for the muffin-tin spheres of Fe,  4d and Ir used in DFT calculations are 2.26 a.u.,  2.31 a.u.  and 2.43 a.u. respectively for the three kind of atoms in each multilayers.  Here we used k-point mesh of 24$\times$24$\times$1 to calculate the magnetic interactions.  Considering the magnetic exchange tensor $\mathrm{J}^{\alpha \beta}_{ij}$ as a [3$\times$3] matrix, the isotropic (Heisenberg) part of the magnetic exchange interactions J$_{ij}$'s and anti-symmetric DMIs D$_{ij}$'s are defined by 
\begin{eqnarray}
\mathrm{\bar{J}}_{ij}= ({\mathrm{J}_{ij}}^{xx}+{\mathrm{J}_{ij}}^{yy}+{\mathrm{J}_{ij}}^{zz})/3,  \nonumber
\\ \nonumber
\mathrm{D}_{ij} = \mid \vec D_{ij} \mid = \sqrt{({\mathrm{D}_{ij}}^x)^2 + ({_{ij}}^y)^2 + ({D_{ij}}^z)^2}, \\
\label{def-int}
\end{eqnarray}
where $\mid \vec D_{ij} \mid $ is the magnitude of the DMI vector.  Here $\alpha$, $\beta$ denotes the Cartesian coordinates (x, y, z), and the $i,j$ denote atomic indices.  We used the convention of positive J$_{ij}$'s as ferromagnetic (FM) and negative J$_{ij}$'s as antiferromagnetic (AFM).  Finally,  we used a spin Hamiltonian using calculated $\bar J_{ij}$ and $\mid \vec D_{ij} \mid $ to investigate the magnetic ground state from both spin dynamics and Monte Carlo simulations without and with external magnetic field \cite{UppASD_book,  uppasd}.

\section{Results and discussion}
\label{sec_III}

\begin{figure*} [ht] 
\centering
\includegraphics[width=1.0\textwidth,angle=0]{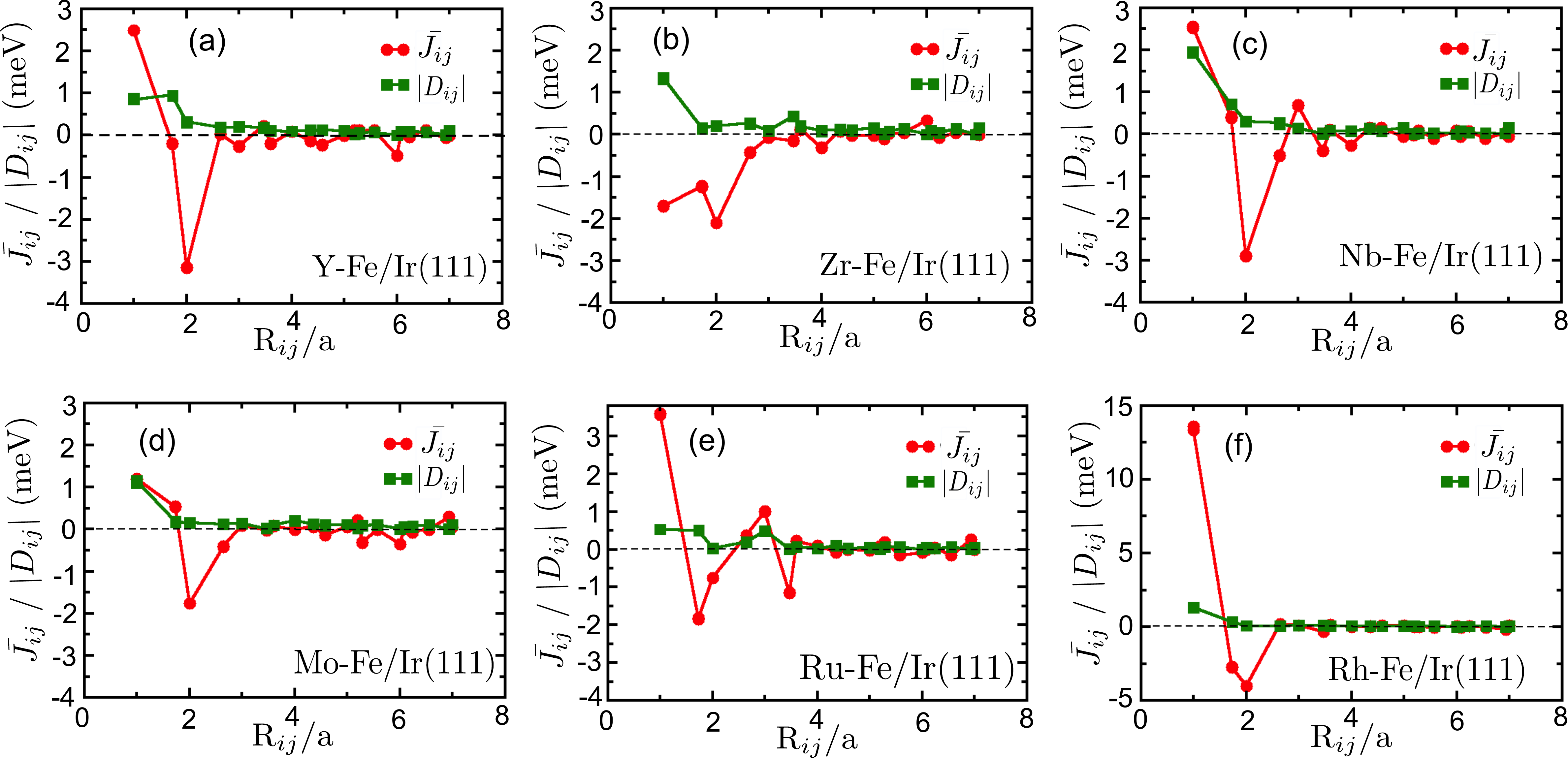} 
\caption{Calculated isotropic exchange interactions $\bar J_{ij}$ and anti-symmetric Dzyaloshinskii-Moriya interactions $\mid \vec D_{ij} \mid $ in 4d/Fe/Ir(111) transition metal multilayers.}
\label{fig1} 
\end{figure*}

\subsection{Tailouring magnetic interactions in 4d/Fe/Ir(111) multilayers}

\par Figure \ref{fig1} (a)-(f) shows the calculated isotropic exchange interactions $\bar J_{ij}$ and interfacial DMI $\mid \vec D_{ij} \mid $   between the Fe atoms as a function of interatomic distances for the different 4d overlayers on Fe/Ir(111) with 4d =  Y,  Zr,  Nb,  Mo, Ru, Rh respectively.  Here positive and negative signs of the isotropic exchange parameters refer to FM and AFM couplings respectively.  Magnetic exchange for the first nearest neighbour (NN) is strongly FM except Zr which has AFM interaction for its first NN Fe-Fe pair.  The values of $\bar J_{1}$  are  presented in table \ref{table1} for a series of 4d transition metals from Y to Rh respectively in 4d/Fe/Ir(111).  {\bl{ $\bar J_{1}$ gradually increases as we move through the 4d series from Y to Ru except Zr and Mo,  and high enough in Rh compared to others 4d transition metals}}.  The magnetic moment of both Fe and 4d transition metals in 4d/Fe/Ir(111) multilayers are also shown in table \ref{table1}.  The magnetic moment of Fe  gradually increases from Y to Rh in 4d/Fe/Ir(111) multilayers except in Zr due to 4d band filling \cite{PhysRevB.97.134405,  PhysRevB.104.024420}.  It induces a small magnetic moments in the adjacent 4d layers leading to the AFM alignment for Mo and Ru,  and FM ordering for others.

  \begin{table*}[ht]
    \small
    \begin{tabular*}{0.97\textwidth}{ p{1.7cm} p{1.7cm} p{1.7cm} p{1.7 cm} p {2.0 cm} p{1.7 cm} p{2.0cm} p{2.0cm} p{1.7cm}}
    \hline\hline
       4d  & $\bar{{J_{1}}}$(meV) & $\frac{\bar{J_1}}{\bar{J_2}}$ & $\frac{\bar{J_1}}{\bar{J_3}}$ & $m_{Fe}$($\mu_B$) & $m_{4d}$($\mu_B$) & $d_{\mathrm{{4d}-{Fe}}}$ (Å) & $d_{\mathrm{{Fe}-{Ir}}}$ (Å) & ${\mathrm{K}}$ (meV)\\
         \hline\hline
    Y : & 2.49 & -11.11 & -0.79 & 2.11 & 0.09 & 1.87 & 1.97 & 0.302  \\
    Zr : &-1.81 & 1.57 & 0.92  &2.38  & 0.22 & 1.89 & 1.98 & 0.929  \\
    Nb : &2.52 & 6.25 & -0.88  &2.21  &0.045 & 1.92 & 2.01 &1.512  \\
    Mo : &1.17 & 2.22 & -0.66  & 2.34  &-0.098  &1.96 & 2.03 &1.022  \\
    Ru : &3.56 & -1.92 & -4.76  &2.42 &-0.22 & 2.03 & 2.07 &1.035  \\
    Rh : &13.57 & -5.0 & -3.45  &2.78  &0.48 & 2.21 & 2.27 &1.136 \\
    Rh \cite{PhysRevLett.120.207201} :  & 9.89 & -2.63 & -4.54  & 2.44  & 0.29 & 1.97 & 2.07 & 1.03 \\
      \hline
    \end{tabular*}
        \caption{ Calculated magnetic exchange interactions for first nearest neighbour (NN) $\bar{{J_{1}}}$,  effective measure of exchange frustration constants $\frac{\bar{J_1}}{\bar{J_2}}$,  $\frac{\bar{J_1}}{\bar{J_3}}$, magnetic moments of $m_{Fe}$,  $m_{4d}$,  relaxed structural parameters and anisotropy energy in fcc stacked 4d/Fe/Ir(111) multilayers.  Here $\bar{{J_{2}}}$ and $\bar{{J_{3}}}$ represent isotropic exchange interactions for second and third NNs respectively.}
    \label{table1}
\end{table*}

\par The trends of variation of magnetic exchange interactions or magnetic moments can be explained by interfacial hybridization of the 3d-Fe layer with both 4d and 5d-Ir.  Interfacial hybridization between 3d-4d metals play a key role in changing the exchange interactions or magnetic moments in the Fe layer as the substrate 5d-Ir is kept fixed here.  We can tune the 3d-4d hybridization in 4d/Fe/Ir(111) with 4d =  Y,  Zr,  Nb,  Mo, Ru, Rh respectively by changing local density of states around the Fermi energy.  As we move through the 4d series from Y to Rh,  the Fermi energy moves from the center to the end of the 4d band.  This shift of the 4d band affects strongly the local density of states of the 3d-Fe layer.   The hybridization of the Fe-3d bands modifies due to the effect of the band filling in the 4d layer.  This increases the magnetic exchange interactions ($\bar J_{1}$) and magnetic moment ($m_{Fe}$) in 4d/Fe/Ir(111) as shown in table \ref{table1}.   The similar trends of varying magnetic moments in Fe have also been observed previously in other 4d/Fe/5d or 5d-Fe/4d multilayers \cite{PhysRevB.104.024420,  PhysRevB.96.094408,  PhysRevB.97.134405}.

\begin{figure} [ht] 
\includegraphics[width=0.5\textwidth,angle=0]{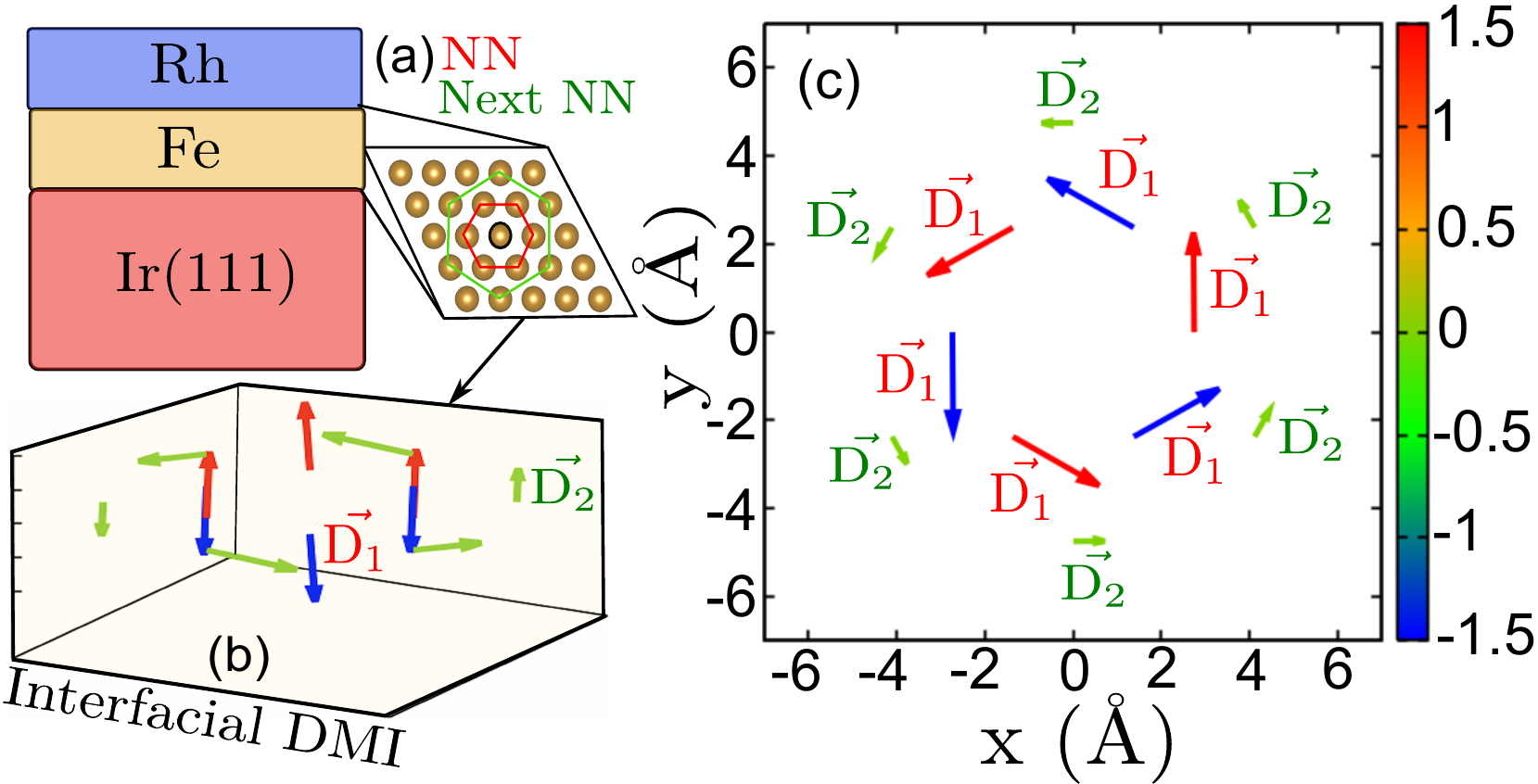} 
\caption{Schematic picture of (a) Rh/Fe/Ir(111) multilayers showing first nearest neighbour (NN) and next NN of Fe atoms coloured by red and green respectively, and (b) interfacial Dzyaloshinskii-Moriya interaction (DMI) vectors in the Fe monolayer. (c) Top view of the same DMI vectors in $xy$ plane.  The colour bar presents the strength of DMI vectors. }
\label{fig2} 
\end{figure}

\begin{figure*} [ht] 
\centering
\includegraphics[width=0.95\textwidth,angle=0]{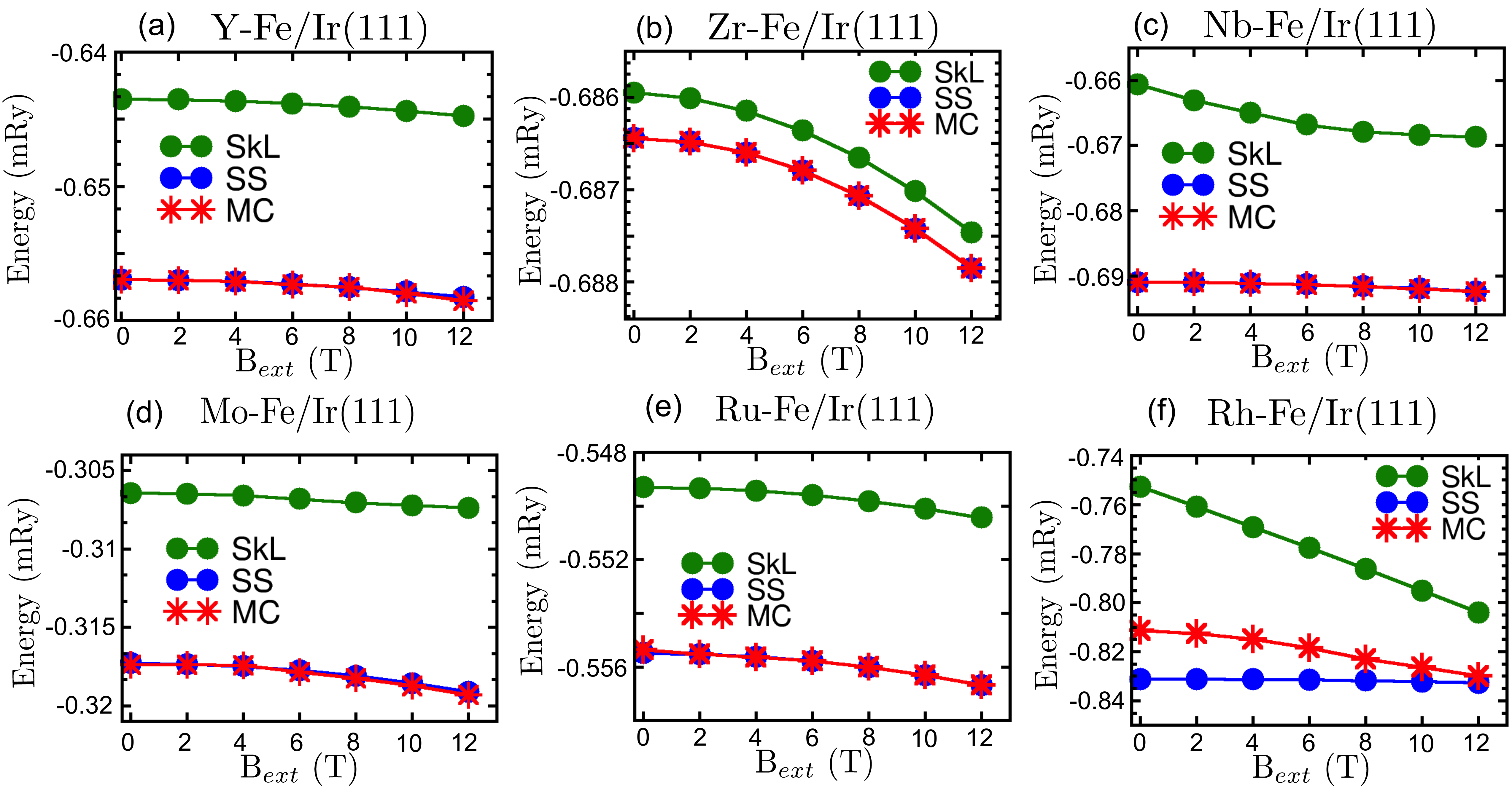} 
\caption{{Calculated energy/spin of spin spiral (SS) and skyrmion lattice (SkL) phase compared to its FM phase {\bl{(i.e E$_{\mathrm{SS/SKL/MC}}$-E$_{\mathrm{FM}}$)}} in fcc stacked 4d/Fe/Ir(111) transition metal multilayers with 4d = Y, Zr, Nb, Mo, Ru and Rh from both spin dynamics and Monte Carlo simulations.  SS phase exists lower in energy for all cases. The combination of SS and SkL appears as a metastable state from Monte Carlo simulations in Rh/Fe/Ir(111). }}
\label{fig3} 
\end{figure*}

\par To get a feeling about on how interfacial coupling of Fe-3d modifies due to the hybridization with 4d transition metals,  we examine the orbital decomposition of magnetic exchange interactions of Fe-3d  orbitals in 4d/Fe/Ir(111) multilayers (see Fig.\ref{sfig1} in appendix) \cite{PhysRevB.105.104418}.  From the multi-orbital analysis,  the total exchange interaction in Fe-3d  can be presented as a sum of three orbital contributions: ${J_{ij}} = {J_{ij}}^{e_{g}-e_{g}} + {J_{ij}}^{t_{2g}-t_{2g}} + {J_{ij}}^{t_{2g}-e_{g}}$ \cite{PhysRevB.105.104418}.  DFT calculations reveal the fact that the ${J_{ij}}^{t_{2g}-e_{g}}$ is mainly responsible for the interfacial coupling effects in 4d/Fe/Ir(111).  The $ {J_{ij}}^{t_{2g}-t_{2g}}$ contribution is FM and ${J_{ij}}^{t_{2g}-e_{g}}$ contribution is AFM in 4d/Fe/Ir(111).  Whereas ${J_{ij}}^{e_{g}-e_{g}}$ has a increasing FM interaction from Zr to Rh except Y 4d/Fe/Ir(111). The isotropic Heisenberg exchange interaction ($\bar{{J_{1}}}$) for NN is AFM in Zr/Fe/Ir(111) due to dominance of AFM ${J_{ij}}^{t_{2g}-e_{g}}$ coupling over FM contributions of $ {J_{ij}}^{t_{2g}-t_{2g}}$ and ${J_{ij}}^{e_{g}-e_{g}}$.  Whereas $\bar{{J_{1}}}$ gets a high FM interaction in Rh/Fe/Ir(111) because of lower contribution of ${J_{ij}}^{t_{2g}-e_{g}}$ compared to other two orbital exchanges.

\par The magnetic exchange interactions in 4d/Fe/Ir(111) multilayers have a strong frustration due to competition between FM NN exchange and AFM for second and third NN exchange interactions.  The ratio of $\frac{\bar{J_1}}{\bar{J_2}}$ and $\frac{\bar{J_1}}{\bar{J_3}}$ measure the effective exchange frustrations in the systems as presented in table \ref{table1}. This frustration originates from the hybridization of the Fe layer with both 4d transition metal and Ir,  resulting {\bl {in a dramatic reduction}} of next NN exchanges compared first NN exchange from Y to Rh in 4d/Fe/Ir(111) transition multilayers.  The  magnetic ground state remains FM considering only NN exchange.  The strong exchange frustration can,  however, results into noncollinear magnetic ground state which will be discussed in the subsequent section.

\begin{figure*} [ht] 
\centering
\includegraphics[width=1.03\textwidth,angle=0]{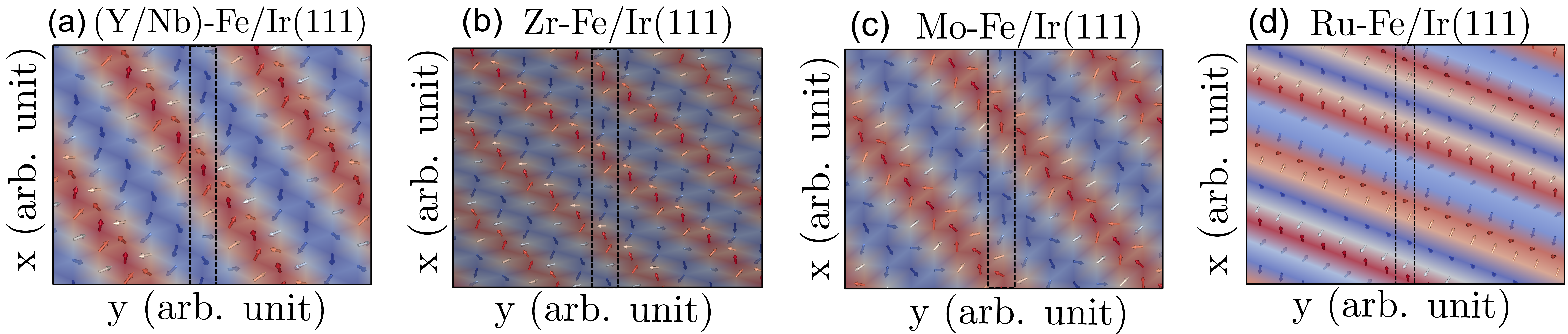} 
\caption{Spin spiral phase in 4d/Fe/Ir(111) multilayers.  Top view of the right rotating spin spirals with a period of (a) 9 atom for Y and Nb, (b) 6 {\bl{ atoms}} for Zr ,(c) 8 atom for Mo  and (d) 4 atom for Ru in 4d/Fe/Ir(111) multilayers.  Rh/Fe/Ir(111) has 6 atom period spin spiral in the same plane. }
\label{fig4} 
\end{figure*}

\par We obtain DMI by inclusion of SOC which leads to spin canting and noncollinear magnetic ground state in 4d/Fe/Ir(111).  Now we can tune DMI in the 4d/Fe/Ir(111) multilayers by considering different 3d/4d interfaces.  Due to the small values of the Heisenberg exchange interactions,  calculated DMI are in similar order of magnitude for Y,  Zr,  Mo,  Ru in 4d/Fe/Ir(111). Whereas it plays a crucial role for the magnetic ground state of Rh/Fe/Ir(111) with inclusion of external magnetic field and will be discussed later.  Figure \ref{fig2} shows the interfacial DMI in Rh/Fe/Ir(111) for NN (D$_1$) and next NN (D$_2$) of Fe atoms.  The size of the vectors in Fig.\ref{fig2}(b)-(c) illustrates on how the DMI strength decreases with distance  and is perpendicular to the bond between interacting Fe atoms.  D$_2$ lies in the plane of Fe monolayer ($xy$-plane) as presented in  Fig.\ref{fig2}(b).  Figure \ref{fig2}(c) represents top view of the same DMI vectors in the $xy$-plane.

\subsection{Spin spiral in 4d/Fe/Ir(111) multilayers}

\par We investigate the magnetic ground state of 4d/Fe/Ir(111) transition metal multilayers from both spin dynamics (SD) and Monte Carlo (MC) simulations.  We use the calculated the isotropic part of Heisenberg exchange interactions J$_{ij}$'s and DMIs D$_{ij}$'s in the spin Hamiltonian below for searching the magnetic ground state.  
\begin{eqnarray}
\begin{aligned}
    {\mathrm{H}}= -\sum_{i,j} {\mathrm{J}}_{ij} {\bm {\mathrm{S}}}_i \cdot {\bm {\mathrm{S}}}_j - \sum_{i,j} {\bm {\mathrm{D}}}_{ij} \cdot ({\bm {\mathrm{S}}}_i \times {\bm {\mathrm{S}}}_j) -{ \sum_i {\mathrm{K}}({S}_{i}^{z})^2} \\ \nonumber - \sum_{i} \bm{B}^{\mathrm{ext}} \cdot \bm{S}_{i},  \nonumber
 \end{aligned}   
\label{eq1}
\end{eqnarray}
where { ${\mathrm{K}}$} is the magnetic anisotropic energy given by E$_c$-E$_p$.  E$_c$ and E$_p$ are the calculated energies when the magnetic quantization axis is along c and in-plane directions respectively.   The last term is the Zeeman term where $\bm{B}^{\mathrm{ext}}$ is the applied external magnetic field.

\par {\bs {The atomistic Landau–Lifshitz–Gilbert (LLG) equation for SD is given by 
\begin{equation}
\label{eq:LLG}
\begin{aligned}
\frac{d \bm{\mathrm S}_i}{d t}=
-\gamma_{\mathrm{L}} \bm{\mathrm S}_i \times\bm{\mathrm B}_i
-\gamma_{\mathrm{L}}\alpha\bm{\mathrm S}_i \times\left(\bm{\mathrm S}_i \times\bm{\mathrm B}_i\right)
\end{aligned}
\end{equation}
This LLG equation is evolved in time until convergence is obtained.
Here, $\bm{\mathrm B}_i = -\partial{\mathrm H}/\partial ( \bm{\mathrm S}_i)$ is the effective field on site $i$ related to the spin Hamiltonian $\mathcal{\mathrm H}$.  The dimensionless (and isotropic) Gilbert damping parameter is here denoted by $\alpha$, while $\gamma_{\mathrm{L}}=\gamma / \left(1+\alpha^2\right)$  is the renormalized gyromagnetic ratio (as a function of the bare one, $\gamma$).  We used a time step of $\Delta t=10^{-16}$\,s to obtain the numerical solution of Eq.\,\eqref{eq:LLG}. We used 150$\times$150$\times$1 supercell in SD simulation with a Gilbert damping factor of 0.1 in Eq.\,\eqref{eq:LLG}. }}

\begin{figure} [t!] 
\includegraphics[width=0.41\textwidth,angle=0]{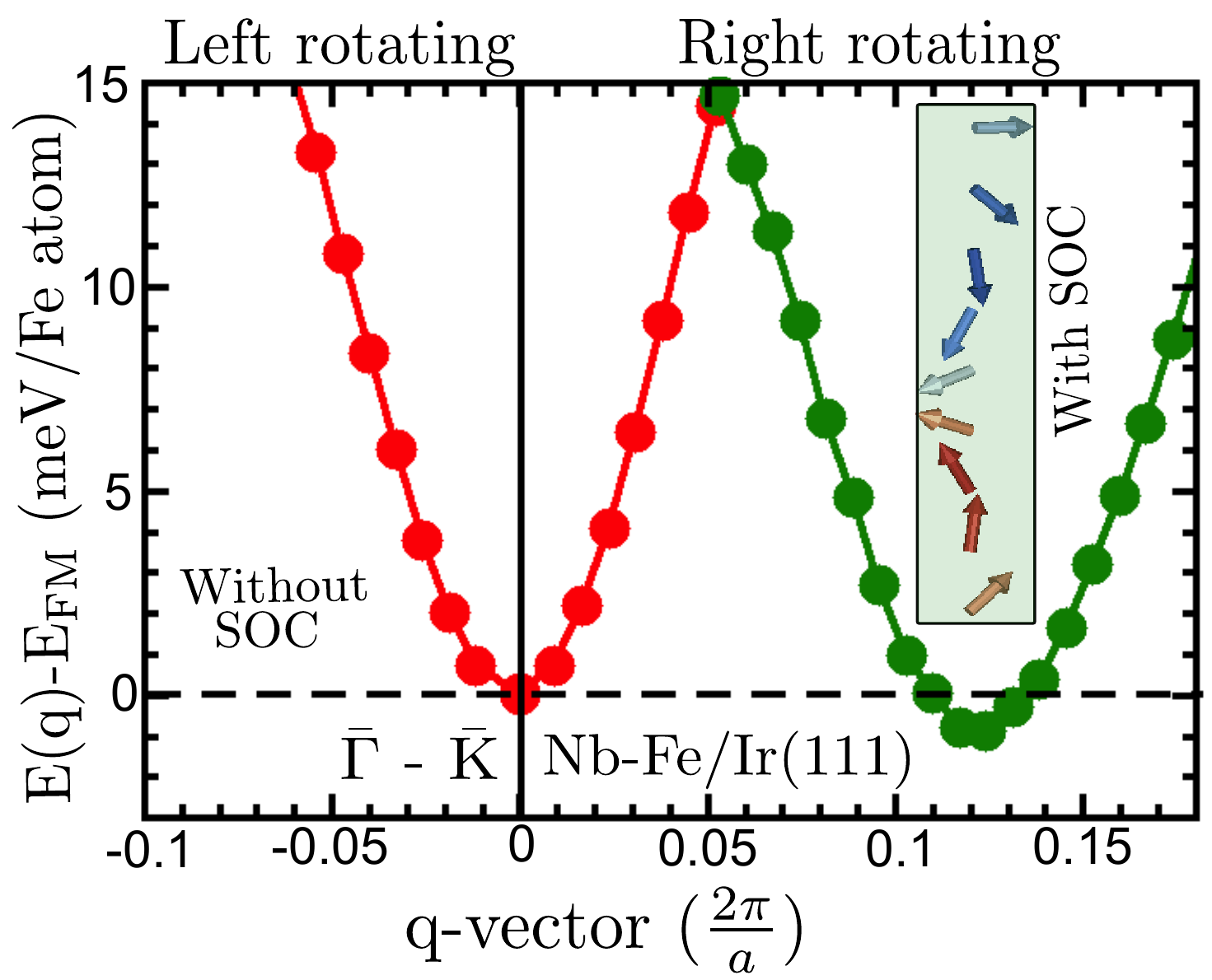} 
\caption{Energy dispersion of spin spirals without and with spin orbit coupling in {\bl {Nb/Fe/Ir(111)}}. }
\label{fig5} 
\end{figure}

\par {\bs{ Our study is also accompanied by heat bath MC simulations  with sufficient temperature annealing for searching the magnetic ground state with external magnetic field \cite{PhysRevB.34.6341}.  The idea behind the heat bath algorithm is to assume that each spin is in contact with a heat bath and is therefore in a local equilibrium with respect to the effective field from all the other spins.  The probability of choosing a state with energy $E$ is proportional to $\exp \left(-E /k_B T\right)$ and normalized so that the total probability summed over all possible states is equal to unity  Here, $k_B$ is the Boltzmann constant, and $T$ is the temperature of the system.  The initial spin configurations were random and annealed down to effectively zero kelvin.   We start from a random spin configuration, and for a given strength of the external magnetic field. Then the system was annealed in 10 steps from $T=500$ K to $T=0.0001$ K.  We used 5$\times 10^{6}$ MC steps at each temperature step to thermalize the system.  In the measurement phase of the MC annealing,  we also used $T=0.0001$ K and 5$\times 10^{6}$ MC steps for searching the magnetic ground state.}}

\par {Figure \ref{fig3}(a)-(f) show the calculated energy of SS and SkL compared to its FM state with $\bm{B}^{\mathrm{ext}}$ for 4d/Fe/Ir(111) with 4d = Y,  Zr,  Nb,  Mo,  Ru and Rh respectively from both SD and MC simulations using the spin Hamiltonian.}  The initial spin texture starts from a random spin configuration in MC simulations.   Corresponding FM phases exist higher in energy compared to SS and SkL phases in 4d/Fe/Ir(111).  We found the magnetic ground state ($\bm{B}^{\mathrm{ext}}$ = 0) of 4d/Fe/Ir(111) is a SS and SkL phase has higher energy compared to SS phase.  The variation of external magnetic field strength has a different effect on the energy of the SS and SkL configurations.  This originates from the out-of-plane spin components of these spin structures entering into the spin Hamiltonian through the Zeeman energy.   SS phase has always lower in energy compared to SkL phase even with external magnetic field for others 4d/Fe/Ir(111) except Rh-Fe/Ir(111).  The energy of the SkL phase in Rh/Fe/Ir(111) gradually decreases with external magnetic field and  has a lower energy value with external magnetic field above 12 T as shown in Fig.\ref{fig3}(f). This indicates of magnetic phase transition from SS to SkL in Rh/Fe/Ir(111) compared to other multilayers which will be discussed subsequently.

\par Figure  \ref{sfig2}(a)-(b) in appendix show the calculated energy from MC simulations with $\bm{B}^{\mathrm{ext}}$ for both  Nb/Fe/Ir(111) and Rh/Fe/Ir(111) respectively.   It gives a $ab$-plane constrained SS phase with a propagation vector ($q_x$,  0, 0) in {\bl{Nb}}/Fe/Ir(111) and it occurs for the others 4d transition metal multilayers also in 4d/Fe/Ir(111) except Rh/Fe/Ir(111).  However,  the mixture of SS and SkL phase appears as meta stable phase in Rh/Fe/Ir(111) as presented in Fig.\ref{sfig2}(b).  The SS phases in 4d/Fe/Ir(111) induced by exchange frustrations persist even with external magnetic field in 4d/Fe/Ir(111) except Rh/Fe/Ir(111).   However,  a moderate amount of exchange frustration leads to SkL phase from SS with external magnetic field in Rh/Fe/Ir(111).

\begin{table}[ht]
    \small
    \begin{tabular*}{0.5\textwidth}{ p{1.5cm} p{3.0cm} p{2.3cm} p{3.0 cm} }
    \hline\hline
       4d/Fe/Ir & q = q$_x\times\frac{2\pi}{a}$ (Å$^{-1}$) & $\lambda=\frac{2\pi}{q}$ (nm) & $\phi=2\pi\cdot{q_x}$ \\
         \hline\hline
    Y : & 0.275 & 2.28 & 43.2$^{\circ}$   \\
    Zr : & 0.367 & 1.71 & 57.6$^{\circ}$   \\
    Nb : & 0.275 & 2.28 & 43.2$^{\circ}$   \\
    Mo : & 0.321 & 1.95 & 50.4$^{\circ}$   \\
    Ru : & 0.642  & 0.97  & 100.8$^{\circ}$    \\
    Rh : & 0.389 &  1.61 & 61.2$^{\circ}$  \\
    Rh \cite{PhysRevLett.120.207201} :  & - & 1.5 & -    \\
      \hline
    \end{tabular*}
        \caption{Calculated spin spiral wave vector ($q$),  wavelength ($\lambda$) and pitch angle ($\phi$) for magnetic {\bl {ground}} state in fcc stacked 4d/Fe/Ir(111) multilayers.}
    \label{table2}
\end{table}

\begin{figure*} [t!] 
\centering
\includegraphics[width=0.99\textwidth,angle=0]{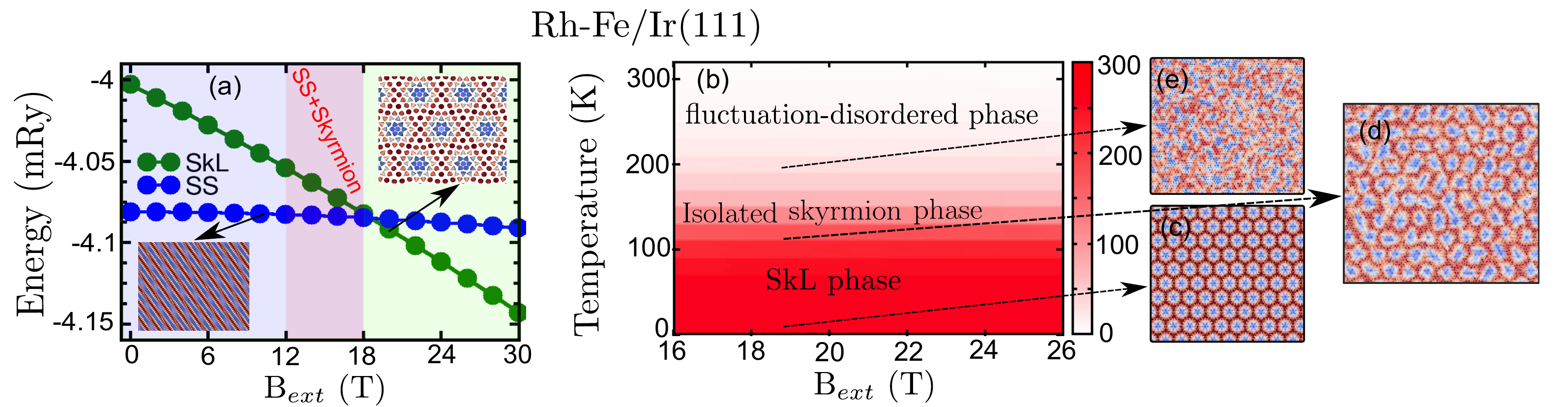} 
\caption{Phase diagram  at (a) T = 0 and (b) T$\neq$ 0 of Rh/Fe/Ir(111) with external magnetic field.  {The calculated energies (energy/spin) for SkL and SS phase in fig(a) are presented compared to corresponding FM state {\bl{(i.e E$_{\mathrm{SS/SKL}}$-E$_{\mathrm{FM}}$)}}.  {\bs{The colour bar in (b) represents normalised skyrmion density.}}}}
\label{fig6} 
\end{figure*}

\par The wave vector ($q$),  wavelength ($\lambda$) and pitch angle ($\phi$) of SS in 4d/Fe/Ir(111) transition multilayers with 4d = Y,  Zr, Nb, Mo, Ru and Rh are displayed in table \ref{table2} corresponding to the SS phase as shown in Fig.\ref{fig4}. { \bs {First, we confirmed the SS phase from both the MC and SD simulations where the initial starting state has started from a random phase.  In the next step, the obtained periodicity is taken as a initial state in SD simulations to find the actual period of SS. }} To investigate the nature SS phase in 4d/Fe/Ir(111), we calculated energy dispersion of SS for {\bl{Nb}}/Fe/Ir(111) without and with SOC along the high symmetry direction $\bar{\Gamma}$-$\bar{K}$ in the two-dimensional Brillouin zone as shown in Fig.\ref{fig5}.  Without SOC,  the left handed and right handed rotating SSs are energetically degenerated.  The energy dispersion has a minimum at the $\bar{\Gamma}$  point which indicates the FM phase of {\bl{Nb}}/Fe/Ir(111) without SOC. The inclusion of SOC induces DMI in 4d/Fe/Ir(111) multilayers due to 3d/4d interface which leads to noncollinear spin structures with canting of the magnetic moments of Fe atoms (as shown in Fig.\ref{fig2}).

\par With the DMI,  SS states with specific rotational sense appears in 4d/Fe/Ir(111).   Figure \ref{fig5} has a minimum energy dispersion at $q_x \approx$ 0.12 in the $\bar{\Gamma}$-$\bar{K}$ direction with SOC  which indicates a right-rotating SS with wave length $\lambda=\frac{2\pi}{q}$ = 2.28 nm in {\bl{Nb}}/Fe/Ir(111).  The energy minima of 1 meV/Fe atom at $q_x \approx$ 0.12 compared to FM phase is attributed due to magnetocrystalline anisotropy with SOC.  However,  the SS wave length tunes due to the hybridization between 3d/4d interface which leads to the different minimum values of q$_{x}$ in energy dispersion for others 4d/Fe/Ir(111) transition multilayers (see  table \ref{table2}).  Such a {\bl {variation}} of SS period due  to temperature effects have been reported in magnetic multilayers of Fe films on Ir(111) using spin-polarized scanning tunneling microscopy \cite{PhysRevLett.119.037202}.  This is attributed due to change of magnetic interactions with temperature effects \cite{PhysRevLett.119.037202}.  However,  the different magnetic interactions, mainly J$_{ij}$'s and D$_{ij}$'s,  for different 4d transition metals in 4d/Fe/Ir(111) are responsible here for different wave lengths, pitch angles and wave vectors of SS in the magnetic ground state of 4d/Fe/Ir(111).  The calculated wave length of SS in Rh/Fe/Ir(111) is $\lambda$ = 1.61 nm which is in good agreement with the experimental report of 1.5 nm  \cite{PhysRevLett.120.207201}.

\subsection{Manupulating skyrmion in 4d/Fe/Ir(111) multilayers}

\par The mixture of SS and skymion has appeared as a meta stable state in fcc stacked Rh/Fe/Ir(111) with a small energy difference of $\approx$ 0.2 meV/Fe atom  without external magnetic field.  Therefore,  it is possible to tune SS into skyrmion phase in Rh/Fe/Ir(111) similar to that of Pd/Fe/Ir(111) \cite{sadhukhan2023spin}.  We perform both SD and MC simulations to determine the phase diagram of fcc stacked  Rh/Fe/Ir(111) in an external magnetic field.  Figure \ref{fig6}(a) presents the simulated phase diagram for  Rh/Fe/Ir(111).  It shows that the SS phase is favourable upto external magnetic field  $\bm{B}^{\mathrm{ext}}$ $\approx$ 12 T and switches into a SkL phase with $\bm{B}^{\mathrm{ext}}$  $\approx$ 18 T.  Rh/Fe/Ir(111) has a extended zone boundary where combination SS and isolated skymion coexist between 12 T $\leq  \bm{B}^{\mathrm{ext}}  \leq$18 T.   {\bs{The phase diagram is also checked replacing the first two terms (isotropic part of Heisenberg exchange interactions J$_{ij}$s and DMI's  D$_{ij}$s) in the spin Hamiltonian by [3x3] magnetic exchange tensor J$_{ij}^{\alpha\beta}$ which confirms that the phase transition points with magnetic fields remain unaffected.}} {\bl{The required magnetic field for Rh/Fe/Ir(111) is high compared to its next 4d transition element i,e  Pd/Fe/Ir(111).  This is due to the fact that the magnetic exchange frustrations are high in Pd/Ir/Ir(111) compared to Rh/Fe/Ir(111) which is also responsible for enhancing skyrmion stability in Pd/Fe/Ir(111) \cite{vonMalottki2017}. }} We also investigate the temperature stability of the SkL phase in Rh/Fe/Ir(111).  {\bs {The thermal stability of the magnetic skyrmion in the Rh/Fe/Ir(111) is investigated using MC simulation with sufficient temperature annealing \cite{PhysRevLett.108.017206,  PhysRevB.88.195137,  PhysRevB.93.024417,  Bottcher_2018}. }} We carefully decrease the system's temperature from 500 K to 0.001 K through 10 steps to study the response of the SkL to thermal agitation for each external  magnetic field between (18-28) T.   The diameter of a skyrmion in {\bl {Rh/Fe/Ir(111)}} is 3.29 nm which is small compared to {\bl {Pd/Fe/Ir(111)}} \cite{PhysRevLett.114.177203,  sadhukhan2023spin} and decreases with the external magnetic field.

\par {\bl{Figure \ref{fig6}(b)-(e)}} show the $B-T$ phase diagram Rh/Fe/Ir(111) from MC simulation.  {\bs{ Our study show that the SkL disappears at a temperature of  T $\leq$ 90 K and forms isolated skyrmions.  The density of isolated skyrmion significantly decreases with increasing temperature T $\geq$ 100 K and enters into fluctuation disordered phase.  The reduction of skyrmion density with decreasing temperature has been reported in a recent work \cite{Kwon2020}.}} The skyrmion phase in Fe/Ir(111) without external magnetic field exist below the temperature T $\approx$ 28.7 K \cite{PhysRevLett.113.077202},  whereas the thermal stability of skyrmion has increased to T $\leq$  90 K with external magnetic field between (18-28) T for Rh/Fe/Ir(111).   Magnetic interactions, in particular,  exchange frustrations and interfacial DMIs discussed above are responsible for increasing the thermal stability of SkL phase in Rh/Fe/Ir(111) compared to Fe/Ir(111).   Our current theoretical investigations provide a deeper insight into the microscopic understanding about it.

Magnetic interactions in 4d/Fe/Ir(111) generates non-collinear spin textures due to a delicate interplay between the symmetric Heisenberg exchange (J$_{ij}$) and interfacial DMI (D$_{ij}$) which generates the spin canting.  J$_{ij}$ increases for Rh compared to other 4d transition metals in 4d/Fe/Ir(111).   Interfacial DMIs D$_{ij}$'s originating from the broken inversion symmetry at the interface of 4d/Fe/Ir(111) multilayers are also changing due to different spin-orbit couplings of 4d transition metal.  The different magnetic interactions along the magnetic anisotropy play the crucial role for the SS ground state in 4d/Fe/Ir(111) with different wave lengths and pitch angles as shown in table \ref{table2}.  Moreover,  Rh/Fe/Ir(111) have a strong exchange frustration due to the hybridization of the Fe-3d layer with both 4d-Rh and Ir-5d layers.  Due to the small values of the Heisenberg exchange interactions J$_{ij}$,  calculated DMIs D$_{ij}$'s are in similar order of magnitude for Y,  Zr,  Mo,  Ru in 4d/Fe/Ir(111) except in Rh/Fe/Ir(111).   The ratio of  $\frac{|{D_{ij}}|}{\bar{J_{ij}}}$ decreases much in  Rh/Fe/Ir(111) compared to others 4d/Fe/Ir(111) which causes the magnetic phase transition in Rh/Fe/Ir(111) with external magnetic field.  Therefore, we can generate SkL phase in Rh/Fe/Ir(111) from SS only in Rh/Fe/Ir(111) compared to others 4d/Fe/Ir(111) multilayers.

\section{Conclusion and outlook}
\label{sec_IV}

\par Magnetic multilayers with noncoplanar magnetic ordering host rich physics.  We have demonstrated from DFT calculations that the isotropic exchange interactions and anti-symmetric DMIs can be tuned at the interface of 4d/Fe/Ir(111) with 4d = Y,  Zr,  Nb,  Mo,  Ru, Rh by modifying the hybridization between 3d-Fe and 4d-transition metal on Ir(111).  Our orbital analysis shows that the exchange interactions between $t_{2g}-e_g$ orbitals play the role for tuning interfacial magnetic coupling in 4d/Fe/Ir(111) multilayers.  We anticipate a strong exchange frustration in 4d/Fe/Ir(111) by tailoring interlayer exchange interactions with 4d = Y,  Zr, Nb,  Mo,  Ru and Rh.  Magnetic exchange frustration increases sufficiently in 4d/Fe/Ir(111) which play crucial role in phase transition from SS to SkL in presence of external magnetic field.  Magnetic ground state of 4d/Fe/Ir(111) transition multilayers is a SS varying wave length between $\approx$ (1-2.5) nm from both SD and MC simulations.  The period of the SS decreases moving through 4d transition metal from Y to Rh.  

\par The exchange frustrations along with interfacial DMIs play an important role in tuning SkL phase from SS phase with external magnetic field.  We observe the magnetic ground state of Rh/Fe/Ir(111) is a SS with spiral period and pitch angle 1.61 nm and 61.2$^{\circ}$ respectively which persists with external magnetic field upto $\approx$ 12 T.  We predict a magnetic phase transition from SS to SkL at external magnetic field $\approx$ 18 T.  In the intermediate values of magnetic field $\approx$ (12-18) T,   we find a crossover region between SS and SkL.  Our atomistic simulations based on DFT parameters suggest that skyrmions in Rh/Fe/Ir(111) is stable at significantly higher temperature upto {\bl{T $\leq$ 90}} K compared to Fe/Ir(111) \cite{PhysRevLett.113.077202}.  We hope that our work will encourage investigations in such magnetic multilayers for engineering skyrmion phase from noncollinear spin texture by providing an important microscopic informations to the experimentalists.  

\section{Acknowledgement}

BS acknowledges Department of Science and Technology, Government of
India, for financial support with reference no DST/WISE-PDF/PM-4/2023 under WISE Post-Doctoral Fellowship programme to carry out this work and,  also acknowledges fruitful discussion with Anna Delin and Anders Bergman.  

\bibliography{xfeir-new}{}

\begin{thebibliography}{68}%
\makeatletter
\providecommand \@ifxundefined [1]{%
 \@ifx{#1\undefined}
}%
\providecommand \@ifnum [1]{%
 \ifnum #1\expandafter \@firstoftwo
 \else \expandafter \@secondoftwo
 \fi
}%
\providecommand \@ifx [1]{%
 \ifx #1\expandafter \@firstoftwo
 \else \expandafter \@secondoftwo
 \fi
}%
\providecommand \natexlab [1]{#1}%
\providecommand \enquote  [1]{``#1''}%
\providecommand \bibnamefont  [1]{#1}%
\providecommand \bibfnamefont [1]{#1}%
\providecommand \citenamefont [1]{#1}%
\providecommand \href@noop [0]{\@secondoftwo}%
\providecommand \href [0]{\begingroup \@sanitize@url \@href}%
\providecommand \@href[1]{\@@startlink{#1}\@@href}%
\providecommand \@@href[1]{\endgroup#1\@@endlink}%
\providecommand \@sanitize@url [0]{\catcode `\\12\catcode `\$12\catcode
  `\&12\catcode `\#12\catcode `\^12\catcode `\_12\catcode `\%12\relax}%
\providecommand \@@startlink[1]{}%
\providecommand \@@endlink[0]{}%
\providecommand \url  [0]{\begingroup\@sanitize@url \@url }%
\providecommand \@url [1]{\endgroup\@href {#1}{\urlprefix }}%
\providecommand \urlprefix  [0]{URL }%
\providecommand \Eprint [0]{\href }%
\providecommand \doibase [0]{http://dx.doi.org/}%
\providecommand \selectlanguage [0]{\@gobble}%
\providecommand \bibinfo  [0]{\@secondoftwo}%
\providecommand \bibfield  [0]{\@secondoftwo}%
\providecommand \translation [1]{[#1]}%
\providecommand \BibitemOpen [0]{}%
\providecommand \bibitemStop [0]{}%
\providecommand \bibitemNoStop [0]{.\EOS\space}%
\providecommand \EOS [0]{\spacefactor3000\relax}%
\providecommand \BibitemShut  [1]{\csname bibitem#1\endcsname}%
\let\auto@bib@innerbib\@empty
\bibitem [{\citenamefont {Heinze}\ \emph {et~al.}(2011)\citenamefont {Heinze},
  \citenamefont {von Bergmann}, \citenamefont {Menzel}, \citenamefont {Brede},
  \citenamefont {Kubetzka}, \citenamefont {Wiesendanger}, \citenamefont
  {Bihlmayer},\ and\ \citenamefont {Bl{\"u}gel}}]{Heinze2011-ke}%
  \BibitemOpen
  \bibfield  {author} {\bibinfo {author} {\bibfnamefont {S}~\bibnamefont
  {Heinze}}, \bibinfo {author} {\bibfnamefont {K}~\bibnamefont {von Bergmann}},
  \bibinfo {author} {\bibfnamefont {M}~\bibnamefont {Menzel}}, \bibinfo
  {author} {\bibfnamefont {J}~\bibnamefont {Brede}}, \bibinfo {author}
  {\bibfnamefont {A}~\bibnamefont {Kubetzka}}, \bibinfo {author} {\bibfnamefont
  {R}~\bibnamefont {Wiesendanger}}, \bibinfo {author} {\bibfnamefont
  {G}~\bibnamefont {Bihlmayer}}, \ and\ \bibinfo {author} {\bibfnamefont
  {S}~\bibnamefont {Bl{\"u}gel}},\ }\bibfield  {title} {\enquote {\bibinfo
  {title} {Spontaneous atomic-scale magnetic skyrmion lattice in two
  dimensions},}\ }\href {https://doi.org/10.1038/nphys2045} {\bibfield
  {journal} {\bibinfo  {journal} {Nature Physics}\ }\textbf {\bibinfo {volume}
  {7}},\ \bibinfo {pages} {713--718} (\bibinfo {year} {2011})}\BibitemShut
  {NoStop}%
\bibitem [{\citenamefont {Romming}\ \emph {et~al.}(2013)\citenamefont
  {Romming}, \citenamefont {Hanneken}, \citenamefont {Menzel}, \citenamefont
  {Bickel}, \citenamefont {Wolter}, \citenamefont {von Bergmann}, \citenamefont
  {Kubetzka},\ and\ \citenamefont {Wiesendanger}}]{Romming2013-uw}%
  \BibitemOpen
  \bibfield  {author} {\bibinfo {author} {\bibfnamefont {N}~\bibnamefont
  {Romming}}, \bibinfo {author} {\bibfnamefont {C}~\bibnamefont {Hanneken}},
  \bibinfo {author} {\bibfnamefont {M}~\bibnamefont {Menzel}}, \bibinfo
  {author} {\bibfnamefont {J~E}\ \bibnamefont {Bickel}}, \bibinfo {author}
  {\bibfnamefont {B}~\bibnamefont {Wolter}}, \bibinfo {author} {\bibfnamefont
  {K}~\bibnamefont {von Bergmann}}, \bibinfo {author} {\bibfnamefont
  {A}~\bibnamefont {Kubetzka}}, \ and\ \bibinfo {author} {\bibfnamefont
  {R}~\bibnamefont {Wiesendanger}},\ }\bibfield  {title} {\enquote {\bibinfo
  {title} {Writing and deleting single magnetic skyrmions},}\ }\href
  {https://doi.org/10.1126/science.1240573} {\bibfield  {journal} {\bibinfo
  {journal} {Science}\ }\textbf {\bibinfo {volume} {341}},\ \bibinfo {pages}
  {636--639} (\bibinfo {year} {2013})}\BibitemShut {NoStop}%
\bibitem [{\citenamefont {Wiesendanger}(2016)}]{Wiesendanger2016-zn}%
  \BibitemOpen
  \bibfield  {author} {\bibinfo {author} {\bibfnamefont {R}~\bibnamefont
  {Wiesendanger}},\ }\bibfield  {title} {\enquote {\bibinfo {title} {Nanoscale
  magnetic skyrmions in metallic films and multilayers: a new twist for
  spintronics},}\ }\href {https://doi.org/10.1038/natrevmats.2016.44}
  {\bibfield  {journal} {\bibinfo  {journal} {Nature Reviews Materials}\
  }\textbf {\bibinfo {volume} {1}},\ \bibinfo {pages} {16044} (\bibinfo {year}
  {2016})}\BibitemShut {NoStop}%
\bibitem [{\citenamefont {Yu}\ \emph {et~al.}(2012)\citenamefont {Yu},
  \citenamefont {Kanazawa}, \citenamefont {Zhang}, \citenamefont {Nagai},
  \citenamefont {Hara}, \citenamefont {Kimoto}, \citenamefont {Matsui},
  \citenamefont {Onose},\ and\ \citenamefont {Tokura}}]{Yu2012-yw}%
  \BibitemOpen
  \bibfield  {author} {\bibinfo {author} {\bibfnamefont {X~Z}\ \bibnamefont
  {Yu}}, \bibinfo {author} {\bibfnamefont {N}~\bibnamefont {Kanazawa}},
  \bibinfo {author} {\bibfnamefont {W~Z}\ \bibnamefont {Zhang}}, \bibinfo
  {author} {\bibfnamefont {T}~\bibnamefont {Nagai}}, \bibinfo {author}
  {\bibfnamefont {T}~\bibnamefont {Hara}}, \bibinfo {author} {\bibfnamefont
  {K}~\bibnamefont {Kimoto}}, \bibinfo {author} {\bibfnamefont {Y}~\bibnamefont
  {Matsui}}, \bibinfo {author} {\bibfnamefont {Y}~\bibnamefont {Onose}}, \ and\
  \bibinfo {author} {\bibfnamefont {Y}~\bibnamefont {Tokura}},\ }\bibfield
  {title} {\enquote {\bibinfo {title} {Skyrmion flow near room temperature in
  an ultralow current density},}\ }\href {https://doi.org/10.1038/ncomms1990}
  {\bibfield  {journal} {\bibinfo  {journal} {Nature Communications}\ }\textbf
  {\bibinfo {volume} {3}},\ \bibinfo {pages} {988} (\bibinfo {year}
  {2012})}\BibitemShut {NoStop}%
\bibitem [{\citenamefont {Fert}\ \emph {et~al.}(2013)\citenamefont {Fert},
  \citenamefont {Cros},\ and\ \citenamefont {Sampaio}}]{Fert2013-wu}%
  \BibitemOpen
  \bibfield  {author} {\bibinfo {author} {\bibfnamefont {A}~\bibnamefont
  {Fert}}, \bibinfo {author} {\bibfnamefont {V}~\bibnamefont {Cros}}, \ and\
  \bibinfo {author} {\bibfnamefont {J}~\bibnamefont {Sampaio}},\ }\bibfield
  {title} {\enquote {\bibinfo {title} {Skyrmions on the track},}\ }\href
  {https://doi.org/10.1038/nnano.2013.29} {\bibfield  {journal} {\bibinfo
  {journal} {Nature Nanotechnology}\ }\textbf {\bibinfo {volume} {8}},\
  \bibinfo {pages} {152--156} (\bibinfo {year} {2013})}\BibitemShut {NoStop}%
\bibitem [{\citenamefont {Nagaosa}\ and\ \citenamefont
  {Tokura}(2013{\natexlab{a}})}]{Nagaosa2013-fx}%
  \BibitemOpen
  \bibfield  {author} {\bibinfo {author} {\bibfnamefont {N}~\bibnamefont
  {Nagaosa}}\ and\ \bibinfo {author} {\bibfnamefont {Y}~\bibnamefont
  {Tokura}},\ }\bibfield  {title} {\enquote {\bibinfo {title} {Topological
  properties and dynamics of magnetic skyrmions},}\ }\href
  {https://doi.org/10.1038/nnano.2013.243} {\bibfield  {journal} {\bibinfo
  {journal} {Nature Nanotechnology}\ }\textbf {\bibinfo {volume} {8}},\
  \bibinfo {pages} {899--911} (\bibinfo {year}
  {2013}{\natexlab{a}})}\BibitemShut {NoStop}%
\bibitem [{\citenamefont {Kiselev}\ \emph {et~al.}(2011)\citenamefont
  {Kiselev}, \citenamefont {Bogdanov}, \citenamefont {Schäfer},\ and\
  \citenamefont {Rößler}}]{Kiselev_2011}%
  \BibitemOpen
  \bibfield  {author} {\bibinfo {author} {\bibfnamefont {N~S}\ \bibnamefont
  {Kiselev}}, \bibinfo {author} {\bibfnamefont {A~N}\ \bibnamefont {Bogdanov}},
  \bibinfo {author} {\bibfnamefont {R}~\bibnamefont {Schäfer}}, \ and\
  \bibinfo {author} {\bibfnamefont {U~K}\ \bibnamefont {Rößler}},\ }\bibfield
   {title} {\enquote {\bibinfo {title} {Chiral skyrmions in thin magnetic
  films: new objects for magnetic storage technologies?}}\ }\href {\doibase
  10.1088/0022-3727/44/39/392001} {\bibfield  {journal} {\bibinfo  {journal}
  {Journal of Physics D: Applied Physics}\ }\textbf {\bibinfo {volume} {44}},\
  \bibinfo {pages} {392001} (\bibinfo {year} {2011})}\BibitemShut {NoStop}%
\bibitem [{\citenamefont {Jonietz}\ \emph {et~al.}(2010)\citenamefont
  {Jonietz}, \citenamefont {M{\"u}hlbauer}, \citenamefont {Pfleiderer},
  \citenamefont {Neubauer}, \citenamefont {M{\"u}nzer}, \citenamefont {Bauer},
  \citenamefont {Adams}, \citenamefont {Georgii}, \citenamefont {B{\"o}ni},
  \citenamefont {Duine}, \citenamefont {Everschor}, \citenamefont {Garst},\
  and\ \citenamefont {Rosch}}]{Jonietz2010-bo}%
  \BibitemOpen
  \bibfield  {author} {\bibinfo {author} {\bibfnamefont {F}~\bibnamefont
  {Jonietz}}, \bibinfo {author} {\bibfnamefont {S}~\bibnamefont
  {M{\"u}hlbauer}}, \bibinfo {author} {\bibfnamefont {C}~\bibnamefont
  {Pfleiderer}}, \bibinfo {author} {\bibfnamefont {A}~\bibnamefont {Neubauer}},
  \bibinfo {author} {\bibfnamefont {W}~\bibnamefont {M{\"u}nzer}}, \bibinfo
  {author} {\bibfnamefont {A}~\bibnamefont {Bauer}}, \bibinfo {author}
  {\bibfnamefont {T}~\bibnamefont {Adams}}, \bibinfo {author} {\bibfnamefont
  {R}~\bibnamefont {Georgii}}, \bibinfo {author} {\bibfnamefont
  {P}~\bibnamefont {B{\"o}ni}}, \bibinfo {author} {\bibfnamefont {R~A}\
  \bibnamefont {Duine}}, \bibinfo {author} {\bibfnamefont {K}~\bibnamefont
  {Everschor}}, \bibinfo {author} {\bibfnamefont {M}~\bibnamefont {Garst}}, \
  and\ \bibinfo {author} {\bibfnamefont {A}~\bibnamefont {Rosch}},\ }\bibfield
  {title} {\enquote {\bibinfo {title} {Spin transfer torques in {MnSi} at
  ultralow current densities},}\ }\href
  {https://doi.org/10.1126/science.1195709} {\bibfield  {journal} {\bibinfo
  {journal} {Science}\ }\textbf {\bibinfo {volume} {330}},\ \bibinfo {pages}
  {1648--1651} (\bibinfo {year} {2010})}\BibitemShut {NoStop}%
\bibitem [{\citenamefont {Schulz}\ \emph {et~al.}(2012)\citenamefont {Schulz},
  \citenamefont {Ritz}, \citenamefont {Bauer}, \citenamefont {Halder},
  \citenamefont {Wagner}, \citenamefont {Franz}, \citenamefont {Pfleiderer},
  \citenamefont {Everschor}, \citenamefont {Garst},\ and\ \citenamefont
  {Rosch}}]{Schulz2012-ax}%
  \BibitemOpen
  \bibfield  {author} {\bibinfo {author} {\bibfnamefont {T}~\bibnamefont
  {Schulz}}, \bibinfo {author} {\bibfnamefont {R}~\bibnamefont {Ritz}},
  \bibinfo {author} {\bibfnamefont {A}~\bibnamefont {Bauer}}, \bibinfo {author}
  {\bibfnamefont {M}~\bibnamefont {Halder}}, \bibinfo {author} {\bibfnamefont
  {M}~\bibnamefont {Wagner}}, \bibinfo {author} {\bibfnamefont {C}~\bibnamefont
  {Franz}}, \bibinfo {author} {\bibfnamefont {C}~\bibnamefont {Pfleiderer}},
  \bibinfo {author} {\bibfnamefont {K}~\bibnamefont {Everschor}}, \bibinfo
  {author} {\bibfnamefont {M}~\bibnamefont {Garst}}, \ and\ \bibinfo {author}
  {\bibfnamefont {A}~\bibnamefont {Rosch}},\ }\bibfield  {title} {\enquote
  {\bibinfo {title} {Emergent electrodynamics of skyrmions in a chiral
  magnet},}\ }\href {https://doi.org/10.1038/nphys2231} {\bibfield  {journal}
  {\bibinfo  {journal} {Nature Physics}\ }\textbf {\bibinfo {volume} {8}},\
  \bibinfo {pages} {301--304} (\bibinfo {year} {2012})}\BibitemShut {NoStop}%
\bibitem [{\citenamefont {Iwasaki}\ \emph
  {et~al.}(2013{\natexlab{a}})\citenamefont {Iwasaki}, \citenamefont
  {Mochizuki},\ and\ \citenamefont {Nagaosa}}]{Iwasaki2013-gi}%
  \BibitemOpen
  \bibfield  {author} {\bibinfo {author} {\bibfnamefont {J}~\bibnamefont
  {Iwasaki}}, \bibinfo {author} {\bibfnamefont {M}~\bibnamefont {Mochizuki}}, \
  and\ \bibinfo {author} {\bibfnamefont {N}~\bibnamefont {Nagaosa}},\
  }\bibfield  {title} {\enquote {\bibinfo {title} {Universal current-velocity
  relation of skyrmion motion in chiral magnets},}\ }\href
  {https://doi.org/10.1038/ncomms2442} {\bibfield  {journal} {\bibinfo
  {journal} {Nature Communications}\ }\textbf {\bibinfo {volume} {4}},\
  \bibinfo {pages} {1463} (\bibinfo {year} {2013}{\natexlab{a}})}\BibitemShut
  {NoStop}%
\bibitem [{\citenamefont {Lin}\ \emph {et~al.}(2013)\citenamefont {Lin},
  \citenamefont {Reichhardt}, \citenamefont {Batista},\ and\ \citenamefont
  {Saxena}}]{PhysRevB.87.214419}%
  \BibitemOpen
  \bibfield  {author} {\bibinfo {author} {\bibfnamefont {S}~\bibnamefont
  {Lin}}, \bibinfo {author} {\bibfnamefont {C}~\bibnamefont {Reichhardt}},
  \bibinfo {author} {\bibfnamefont {C~D.}\ \bibnamefont {Batista}}, \ and\
  \bibinfo {author} {\bibfnamefont {A}~\bibnamefont {Saxena}},\ }\bibfield
  {title} {\enquote {\bibinfo {title} {Particle model for skyrmions in metallic
  chiral magnets: Dynamics, pinning, and creep},}\ }\href {\doibase
  10.1103/PhysRevB.87.214419} {\bibfield  {journal} {\bibinfo  {journal} {Phys.
  Rev. B}\ }\textbf {\bibinfo {volume} {87}},\ \bibinfo {pages} {214419}
  (\bibinfo {year} {2013})}\BibitemShut {NoStop}%
\bibitem [{\citenamefont {Iwasaki}\ \emph
  {et~al.}(2013{\natexlab{b}})\citenamefont {Iwasaki}, \citenamefont
  {Mochizuki},\ and\ \citenamefont {Nagaosa}}]{Iwasaki2013-fg}%
  \BibitemOpen
  \bibfield  {author} {\bibinfo {author} {\bibfnamefont {J}~\bibnamefont
  {Iwasaki}}, \bibinfo {author} {\bibfnamefont {M}~\bibnamefont {Mochizuki}}, \
  and\ \bibinfo {author} {\bibfnamefont {N}~\bibnamefont {Nagaosa}},\
  }\bibfield  {title} {\enquote {\bibinfo {title} {Current-induced skyrmion
  dynamics in constricted geometries},}\ }\href
  {https://doi.org/10.1038/nnano.2013.176} {\bibfield  {journal} {\bibinfo
  {journal} {Nature Nanotechnology}\ }\textbf {\bibinfo {volume} {8}},\
  \bibinfo {pages} {742--747} (\bibinfo {year}
  {2013}{\natexlab{b}})}\BibitemShut {NoStop}%
\bibitem [{\citenamefont {Zhang}\ \emph {et~al.}(2015)\citenamefont {Zhang},
  \citenamefont {Zhao}, \citenamefont {Fangohr}, \citenamefont {Liu},
  \citenamefont {Xia}, \citenamefont {Xia},\ and\ \citenamefont
  {Morvan}}]{Zhang2015-mn}%
  \BibitemOpen
  \bibfield  {author} {\bibinfo {author} {\bibfnamefont {X}~\bibnamefont
  {Zhang}}, \bibinfo {author} {\bibfnamefont {G~P}\ \bibnamefont {Zhao}},
  \bibinfo {author} {\bibfnamefont {H}~\bibnamefont {Fangohr}}, \bibinfo
  {author} {\bibfnamefont {J~P}\ \bibnamefont {Liu}}, \bibinfo {author}
  {\bibfnamefont {W~X}\ \bibnamefont {Xia}}, \bibinfo {author} {\bibfnamefont
  {J}~\bibnamefont {Xia}}, \ and\ \bibinfo {author} {\bibfnamefont {F~J}\
  \bibnamefont {Morvan}},\ }\bibfield  {title} {\enquote {\bibinfo {title}
  {Skyrmion-skyrmion and skyrmion-edge repulsions in skyrmion-based racetrack
  memory},}\ }\href {https://doi.org/10.1038/srep07643} {\bibfield  {journal}
  {\bibinfo  {journal} {Scientific Reports}\ }\textbf {\bibinfo {volume} {5}},\
  \bibinfo {pages} {7643} (\bibinfo {year} {2015})}\BibitemShut {NoStop}%
\bibitem [{\citenamefont {Shu}\ \emph {et~al.}(2022)\citenamefont {Shu},
  \citenamefont {Li}, \citenamefont {Xia}, \citenamefont {Lai}, \citenamefont
  {Hou}, \citenamefont {Zhao}, \citenamefont {Zhang}, \citenamefont {Zhou},
  \citenamefont {Liu},\ and\ \citenamefont {Zhao}}]{shu2022realization}%
  \BibitemOpen
  \bibfield  {author} {\bibinfo {author} {\bibfnamefont {Y}~\bibnamefont
  {Shu}}, \bibinfo {author} {\bibfnamefont {Q}~\bibnamefont {Li}}, \bibinfo
  {author} {\bibfnamefont {J}~\bibnamefont {Xia}}, \bibinfo {author}
  {\bibfnamefont {P}~\bibnamefont {Lai}}, \bibinfo {author} {\bibfnamefont
  {Z}~\bibnamefont {Hou}}, \bibinfo {author} {\bibfnamefont {Y}~\bibnamefont
  {Zhao}}, \bibinfo {author} {\bibfnamefont {D}~\bibnamefont {Zhang}}, \bibinfo
  {author} {\bibfnamefont {Y}~\bibnamefont {Zhou}}, \bibinfo {author}
  {\bibfnamefont {X}~\bibnamefont {Liu}}, \ and\ \bibinfo {author}
  {\bibfnamefont {G}~\bibnamefont {Zhao}},\ }\bibfield  {title} {\enquote
  {\bibinfo {title} {Realization of the skyrmionic logic gates and diodes in
  the same racetrack with enhanced and modified edges},}\ }\href
  {https://doi.org/10.1063/5.0097152} {\bibfield  {journal} {\bibinfo
  {journal} {Applied Physics Letters}\ }\textbf {\bibinfo {volume} {121}}
  (\bibinfo {year} {2022})}\BibitemShut {NoStop}%
\bibitem [{\citenamefont {Shen}\ \emph {et~al.}(2018)\citenamefont {Shen},
  \citenamefont {Xia}, \citenamefont {Zhao}, \citenamefont {Zhang},
  \citenamefont {Ezawa}, \citenamefont {Tretiakov}, \citenamefont {Liu},\ and\
  \citenamefont {Zhou}}]{PhysRevB.98.134448}%
  \BibitemOpen
  \bibfield  {author} {\bibinfo {author} {\bibfnamefont {L}~\bibnamefont
  {Shen}}, \bibinfo {author} {\bibfnamefont {J}~\bibnamefont {Xia}}, \bibinfo
  {author} {\bibfnamefont {G}~\bibnamefont {Zhao}}, \bibinfo {author}
  {\bibfnamefont {X}~\bibnamefont {Zhang}}, \bibinfo {author} {\bibfnamefont
  {M}~\bibnamefont {Ezawa}}, \bibinfo {author} {\bibfnamefont {O~A.}\
  \bibnamefont {Tretiakov}}, \bibinfo {author} {\bibfnamefont {X}~\bibnamefont
  {Liu}}, \ and\ \bibinfo {author} {\bibfnamefont {Y}~\bibnamefont {Zhou}},\
  }\bibfield  {title} {\enquote {\bibinfo {title} {Dynamics of the
  antiferromagnetic skyrmion induced by a magnetic anisotropy gradient},}\
  }\href {\doibase 10.1103/PhysRevB.98.134448} {\bibfield  {journal} {\bibinfo
  {journal} {Phys. Rev. B}\ }\textbf {\bibinfo {volume} {98}},\ \bibinfo
  {pages} {134448} (\bibinfo {year} {2018})}\BibitemShut {NoStop}%
\bibitem [{\citenamefont {Nagaosa}\ and\ \citenamefont
  {Tokura}(2013{\natexlab{b}})}]{Nagaosa2013-ns}%
  \BibitemOpen
  \bibfield  {author} {\bibinfo {author} {\bibfnamefont {N}~\bibnamefont
  {Nagaosa}}\ and\ \bibinfo {author} {\bibfnamefont {Y}~\bibnamefont
  {Tokura}},\ }\bibfield  {title} {\enquote {\bibinfo {title} {Topological
  properties and dynamics of magnetic skyrmions},}\ }\href
  {https://doi.org/10.1038/nnano.2013.243} {\bibfield  {journal} {\bibinfo
  {journal} {Nature Nanotechnology}\ }\textbf {\bibinfo {volume} {8}},\
  \bibinfo {pages} {899--911} (\bibinfo {year}
  {2013}{\natexlab{b}})}\BibitemShut {NoStop}%
\bibitem [{\citenamefont {M{\"u}hlbauer}\ \emph {et~al.}(2009)\citenamefont
  {M{\"u}hlbauer}, \citenamefont {Binz}, \citenamefont {Jonietz}, \citenamefont
  {Pfleiderer}, \citenamefont {Rosch}, \citenamefont {Neubauer}, \citenamefont
  {Georgii},\ and\ \citenamefont {B{\"o}ni}}]{Muhlbauer2009-hu}%
  \BibitemOpen
  \bibfield  {author} {\bibinfo {author} {\bibfnamefont {S}~\bibnamefont
  {M{\"u}hlbauer}}, \bibinfo {author} {\bibfnamefont {B}~\bibnamefont {Binz}},
  \bibinfo {author} {\bibfnamefont {F}~\bibnamefont {Jonietz}}, \bibinfo
  {author} {\bibfnamefont {C}~\bibnamefont {Pfleiderer}}, \bibinfo {author}
  {\bibfnamefont {A}~\bibnamefont {Rosch}}, \bibinfo {author} {\bibfnamefont
  {A}~\bibnamefont {Neubauer}}, \bibinfo {author} {\bibfnamefont
  {R}~\bibnamefont {Georgii}}, \ and\ \bibinfo {author} {\bibfnamefont
  {P}~\bibnamefont {B{\"o}ni}},\ }\bibfield  {title} {\enquote {\bibinfo
  {title} {Skyrmion lattice in a chiral magnet},}\ }\href
  {https://doi.org/10.1126/science.1166767} {\bibfield  {journal} {\bibinfo
  {journal} {Science}\ }\textbf {\bibinfo {volume} {323}},\ \bibinfo {pages}
  {915--919} (\bibinfo {year} {2009})}\BibitemShut {NoStop}%
\bibitem [{\citenamefont {Yu}\ \emph {et~al.}(2010)\citenamefont {Yu},
  \citenamefont {Onose}, \citenamefont {Kanazawa}, \citenamefont {Park},
  \citenamefont {Han}, \citenamefont {Matsui}, \citenamefont {Nagaosa},\ and\
  \citenamefont {Tokura}}]{Yu2010-ea}%
  \BibitemOpen
  \bibfield  {author} {\bibinfo {author} {\bibfnamefont {X~Z}\ \bibnamefont
  {Yu}}, \bibinfo {author} {\bibfnamefont {Y}~\bibnamefont {Onose}}, \bibinfo
  {author} {\bibfnamefont {N}~\bibnamefont {Kanazawa}}, \bibinfo {author}
  {\bibfnamefont {J~H}\ \bibnamefont {Park}}, \bibinfo {author} {\bibfnamefont
  {J~H}\ \bibnamefont {Han}}, \bibinfo {author} {\bibfnamefont {Y}~\bibnamefont
  {Matsui}}, \bibinfo {author} {\bibfnamefont {N}~\bibnamefont {Nagaosa}}, \
  and\ \bibinfo {author} {\bibfnamefont {Y}~\bibnamefont {Tokura}},\ }\bibfield
   {title} {\enquote {\bibinfo {title} {Real-space observation of a
  two-dimensional skyrmion crystal},}\ }\href
  {https://doi.org/10.1038/nature09124} {\bibfield  {journal} {\bibinfo
  {journal} {Nature}\ }\textbf {\bibinfo {volume} {465}},\ \bibinfo {pages}
  {901--904} (\bibinfo {year} {2010})}\BibitemShut {NoStop}%
\bibitem [{\citenamefont {Wilhelm}\ \emph {et~al.}(2011)\citenamefont
  {Wilhelm}, \citenamefont {Baenitz}, \citenamefont {Schmidt}, \citenamefont
  {R\"o\ss{}ler}, \citenamefont {Leonov},\ and\ \citenamefont
  {Bogdanov}}]{PhysRevLett.107.127203}%
  \BibitemOpen
  \bibfield  {author} {\bibinfo {author} {\bibfnamefont {H.}~\bibnamefont
  {Wilhelm}}, \bibinfo {author} {\bibfnamefont {M.}~\bibnamefont {Baenitz}},
  \bibinfo {author} {\bibfnamefont {M.}~\bibnamefont {Schmidt}}, \bibinfo
  {author} {\bibfnamefont {U.~K.}\ \bibnamefont {R\"o\ss{}ler}}, \bibinfo
  {author} {\bibfnamefont {A.~A.}\ \bibnamefont {Leonov}}, \ and\ \bibinfo
  {author} {\bibfnamefont {A.~N.}\ \bibnamefont {Bogdanov}},\ }\bibfield
  {title} {\enquote {\bibinfo {title} {Precursor phenomena at the magnetic
  ordering of the cubic helimagnet fege},}\ }\href {\doibase
  10.1103/PhysRevLett.107.127203} {\bibfield  {journal} {\bibinfo  {journal}
  {Phys. Rev. Lett.}\ }\textbf {\bibinfo {volume} {107}},\ \bibinfo {pages}
  {127203} (\bibinfo {year} {2011})}\BibitemShut {NoStop}%
\bibitem [{\citenamefont {M\"unzer}\ \emph {et~al.}(2010)\citenamefont
  {M\"unzer}, \citenamefont {Neubauer}, \citenamefont {Adams}, \citenamefont
  {M\"uhlbauer}, \citenamefont {Franz}, \citenamefont {Jonietz}, \citenamefont
  {Georgii}, \citenamefont {B\"oni}, \citenamefont {Pedersen}, \citenamefont
  {Schmidt}, \citenamefont {Rosch},\ and\ \citenamefont
  {Pfleiderer}}]{PhysRevB.81.041203}%
  \BibitemOpen
  \bibfield  {author} {\bibinfo {author} {\bibfnamefont {W.}~\bibnamefont
  {M\"unzer}}, \bibinfo {author} {\bibfnamefont {A.}~\bibnamefont {Neubauer}},
  \bibinfo {author} {\bibfnamefont {T.}~\bibnamefont {Adams}}, \bibinfo
  {author} {\bibfnamefont {S.}~\bibnamefont {M\"uhlbauer}}, \bibinfo {author}
  {\bibfnamefont {C.}~\bibnamefont {Franz}}, \bibinfo {author} {\bibfnamefont
  {F.}~\bibnamefont {Jonietz}}, \bibinfo {author} {\bibfnamefont
  {R.}~\bibnamefont {Georgii}}, \bibinfo {author} {\bibfnamefont
  {P.}~\bibnamefont {B\"oni}}, \bibinfo {author} {\bibfnamefont
  {B.}~\bibnamefont {Pedersen}}, \bibinfo {author} {\bibfnamefont
  {M.}~\bibnamefont {Schmidt}}, \bibinfo {author} {\bibfnamefont
  {A.}~\bibnamefont {Rosch}}, \ and\ \bibinfo {author} {\bibfnamefont
  {C.}~\bibnamefont {Pfleiderer}},\ }\bibfield  {title} {\enquote {\bibinfo
  {title} {Skyrmion lattice in the doped semiconductor
  ${\text{fe}}_{1\ensuremath{-}x}{\text{co}}_{x}\text{Si}$},}\ }\href {\doibase
  10.1103/PhysRevB.81.041203} {\bibfield  {journal} {\bibinfo  {journal} {Phys.
  Rev. B}\ }\textbf {\bibinfo {volume} {81}},\ \bibinfo {pages} {041203}
  (\bibinfo {year} {2010})}\BibitemShut {NoStop}%
\bibitem [{\citenamefont {Tonomura}\ \emph {et~al.}(2012)\citenamefont
  {Tonomura}, \citenamefont {Yu}, \citenamefont {Yanagisawa}, \citenamefont
  {Matsuda}, \citenamefont {Onose}, \citenamefont {Kanazawa}, \citenamefont
  {Park},\ and\ \citenamefont {Tokura}}]{Tonomura2012-mp}%
  \BibitemOpen
  \bibfield  {author} {\bibinfo {author} {\bibfnamefont {A}~\bibnamefont
  {Tonomura}}, \bibinfo {author} {\bibfnamefont {X}~\bibnamefont {Yu}},
  \bibinfo {author} {\bibfnamefont {K}~\bibnamefont {Yanagisawa}}, \bibinfo
  {author} {\bibfnamefont {T}~\bibnamefont {Matsuda}}, \bibinfo {author}
  {\bibfnamefont {Y}~\bibnamefont {Onose}}, \bibinfo {author} {\bibfnamefont
  {N}~\bibnamefont {Kanazawa}}, \bibinfo {author} {\bibfnamefont {H~S}\
  \bibnamefont {Park}}, \ and\ \bibinfo {author} {\bibfnamefont
  {Y}~\bibnamefont {Tokura}},\ }\bibfield  {title} {\enquote {\bibinfo {title}
  {{Real-Space} observation of skyrmion lattice in helimagnet {MnSi} thin
  samples},}\ }\href {https://doi.org/10.1021/nl300073m} {\bibfield  {journal}
  {\bibinfo  {journal} {Nano Lett.}\ }\textbf {\bibinfo {volume} {12}},\
  \bibinfo {pages} {1673--1677} (\bibinfo {year} {2012})}\BibitemShut {NoStop}%
\bibitem [{\citenamefont {Yu}\ \emph {et~al.}(2011)\citenamefont {Yu},
  \citenamefont {Kanazawa}, \citenamefont {Onose}, \citenamefont {Kimoto},
  \citenamefont {Zhang}, \citenamefont {Ishiwata}, \citenamefont {Matsui},\
  and\ \citenamefont {Tokura}}]{Yu2011-ix}%
  \BibitemOpen
  \bibfield  {author} {\bibinfo {author} {\bibfnamefont {X~Z}\ \bibnamefont
  {Yu}}, \bibinfo {author} {\bibfnamefont {N}~\bibnamefont {Kanazawa}},
  \bibinfo {author} {\bibfnamefont {Y}~\bibnamefont {Onose}}, \bibinfo {author}
  {\bibfnamefont {K}~\bibnamefont {Kimoto}}, \bibinfo {author} {\bibfnamefont
  {W~Z}\ \bibnamefont {Zhang}}, \bibinfo {author} {\bibfnamefont
  {S}~\bibnamefont {Ishiwata}}, \bibinfo {author} {\bibfnamefont
  {Y}~\bibnamefont {Matsui}}, \ and\ \bibinfo {author} {\bibfnamefont
  {Y}~\bibnamefont {Tokura}},\ }\bibfield  {title} {\enquote {\bibinfo {title}
  {Near room-temperature formation of a skyrmion crystal in thin-films of the
  helimagnet {FeGe}},}\ }\href {https://doi.org/10.1038/nmat2916} {\bibfield
  {journal} {\bibinfo  {journal} {Nature Materials}\ }\textbf {\bibinfo
  {volume} {10}},\ \bibinfo {pages} {106--109} (\bibinfo {year}
  {2011})}\BibitemShut {NoStop}%
\bibitem [{\citenamefont {Zhang}\ \emph {et~al.}(2020)\citenamefont {Zhang},
  \citenamefont {Xu}, \citenamefont {Chen}, \citenamefont {Nahas},
  \citenamefont {Prokhorenko},\ and\ \citenamefont
  {Bellaiche}}]{PhysRevB.102.241107}%
  \BibitemOpen
  \bibfield  {author} {\bibinfo {author} {\bibfnamefont {Y}~\bibnamefont
  {Zhang}}, \bibinfo {author} {\bibfnamefont {C}~\bibnamefont {Xu}}, \bibinfo
  {author} {\bibfnamefont {P}~\bibnamefont {Chen}}, \bibinfo {author}
  {\bibfnamefont {Y}~\bibnamefont {Nahas}}, \bibinfo {author} {\bibfnamefont
  {S}~\bibnamefont {Prokhorenko}}, \ and\ \bibinfo {author} {\bibfnamefont
  {L}~\bibnamefont {Bellaiche}},\ }\bibfield  {title} {\enquote {\bibinfo
  {title} {Emergence of skyrmionium in a two-dimensional
  ${\mathrm{crge}(\mathrm{se},\mathrm{te})}_{3}$ janus monolayer},}\ }\href
  {\doibase 10.1103/PhysRevB.102.241107} {\bibfield  {journal} {\bibinfo
  {journal} {Phys. Rev. B}\ }\textbf {\bibinfo {volume} {102}},\ \bibinfo
  {pages} {241107} (\bibinfo {year} {2020})}\BibitemShut {NoStop}%
\bibitem [{\citenamefont {Li}\ \emph {et~al.}(2022)\citenamefont {Li},
  \citenamefont {Haldar},\ and\ \citenamefont {Heinze}}]{Li2022-tx}%
  \BibitemOpen
  \bibfield  {author} {\bibinfo {author} {\bibfnamefont {D}~\bibnamefont {Li}},
  \bibinfo {author} {\bibfnamefont {S}~\bibnamefont {Haldar}}, \ and\ \bibinfo
  {author} {\bibfnamefont {S}~\bibnamefont {Heinze}},\ }\bibfield  {title}
  {\enquote {\bibinfo {title} {{Strain-Driven} {Zero-Field} near-10 nm
  skyrmions in {Two-Dimensional} van der waals heterostructures},}\ }\href
  {https://doi.org/10.1021/acs.nanolett.2c03287} {\bibfield  {journal}
  {\bibinfo  {journal} {Nano Lett.}\ }\textbf {\bibinfo {volume} {22}},\
  \bibinfo {pages} {7706--7713} (\bibinfo {year} {2022})}\BibitemShut {NoStop}%
\bibitem [{\citenamefont {Li}\ \emph {et~al.}(2023)\citenamefont {Li},
  \citenamefont {Yu}, \citenamefont {Liang}, \citenamefont {Ga},\ and\
  \citenamefont {Yang}}]{PhysRevB.107.054408}%
  \BibitemOpen
  \bibfield  {author} {\bibinfo {author} {\bibfnamefont {P}~\bibnamefont {Li}},
  \bibinfo {author} {\bibfnamefont {D}~\bibnamefont {Yu}}, \bibinfo {author}
  {\bibfnamefont {J}~\bibnamefont {Liang}}, \bibinfo {author} {\bibfnamefont
  {Y}~\bibnamefont {Ga}}, \ and\ \bibinfo {author} {\bibfnamefont
  {H}~\bibnamefont {Yang}},\ }\bibfield  {title} {\enquote {\bibinfo {title}
  {Topological spin textures in $1t$-phase janus magnets: Interplay between
  dzyaloshinskii-moriya interaction, magnetic frustration, and isotropic
  higher-order interactions},}\ }\href {\doibase 10.1103/PhysRevB.107.054408}
  {\bibfield  {journal} {\bibinfo  {journal} {Phys. Rev. B}\ }\textbf {\bibinfo
  {volume} {107}},\ \bibinfo {pages} {054408} (\bibinfo {year}
  {2023})}\BibitemShut {NoStop}%
\bibitem [{\citenamefont {Bode}\ \emph {et~al.}(2007)\citenamefont {Bode},
  \citenamefont {Heide}, \citenamefont {von Bergmann}, \citenamefont
  {Ferriani}, \citenamefont {Heinze}, \citenamefont {Bihlmayer}, \citenamefont
  {Kubetzka}, \citenamefont {Pietzsch}, \citenamefont {Bl{\"u}gel},\ and\
  \citenamefont {Wiesendanger}}]{Bode2007-ku}%
  \BibitemOpen
  \bibfield  {author} {\bibinfo {author} {\bibfnamefont {M}~\bibnamefont
  {Bode}}, \bibinfo {author} {\bibfnamefont {M}~\bibnamefont {Heide}}, \bibinfo
  {author} {\bibfnamefont {K}~\bibnamefont {von Bergmann}}, \bibinfo {author}
  {\bibfnamefont {P}~\bibnamefont {Ferriani}}, \bibinfo {author} {\bibfnamefont
  {S}~\bibnamefont {Heinze}}, \bibinfo {author} {\bibfnamefont {G}~\bibnamefont
  {Bihlmayer}}, \bibinfo {author} {\bibfnamefont {A}~\bibnamefont {Kubetzka}},
  \bibinfo {author} {\bibfnamefont {O}~\bibnamefont {Pietzsch}}, \bibinfo
  {author} {\bibfnamefont {S}~\bibnamefont {Bl{\"u}gel}}, \ and\ \bibinfo
  {author} {\bibfnamefont {R}~\bibnamefont {Wiesendanger}},\ }\bibfield
  {title} {\enquote {\bibinfo {title} {Chiral magnetic order at surfaces driven
  by inversion asymmetry},}\ }\href {https://doi.org/10.1038/nature05802}
  {\bibfield  {journal} {\bibinfo  {journal} {Nature}\ }\textbf {\bibinfo
  {volume} {447}},\ \bibinfo {pages} {190--193} (\bibinfo {year}
  {2007})}\BibitemShut {NoStop}%
\bibitem [{\citenamefont {Simon}\ \emph {et~al.}(2018)\citenamefont {Simon},
  \citenamefont {R\'ozsa}, \citenamefont {Palot\'as},\ and\ \citenamefont
  {Szunyogh}}]{PhysRevB.97.134405}%
  \BibitemOpen
  \bibfield  {author} {\bibinfo {author} {\bibfnamefont {E.}~\bibnamefont
  {Simon}}, \bibinfo {author} {\bibfnamefont {L.}~\bibnamefont {R\'ozsa}},
  \bibinfo {author} {\bibfnamefont {K.}~\bibnamefont {Palot\'as}}, \ and\
  \bibinfo {author} {\bibfnamefont {L.}~\bibnamefont {Szunyogh}},\ }\bibfield
  {title} {\enquote {\bibinfo {title} {Magnetism of a co monolayer on pt(111)
  capped by overlayers of $5d$ elements: A spin-model study},}\ }\href
  {\doibase 10.1103/PhysRevB.97.134405} {\bibfield  {journal} {\bibinfo
  {journal} {Phys. Rev. B}\ }\textbf {\bibinfo {volume} {97}},\ \bibinfo
  {pages} {134405} (\bibinfo {year} {2018})}\BibitemShut {NoStop}%
\bibitem [{\citenamefont {Meyer}\ \emph {et~al.}(2017)\citenamefont {Meyer},
  \citenamefont {Dup\'e}, \citenamefont {Ferriani},\ and\ \citenamefont
  {Heinze}}]{PhysRevB.96.094408}%
  \BibitemOpen
  \bibfield  {author} {\bibinfo {author} {\bibfnamefont {S}~\bibnamefont
  {Meyer}}, \bibinfo {author} {\bibfnamefont {B}~\bibnamefont {Dup\'e}},
  \bibinfo {author} {\bibfnamefont {P}~\bibnamefont {Ferriani}}, \ and\
  \bibinfo {author} {\bibfnamefont {S}~\bibnamefont {Heinze}},\ }\bibfield
  {title} {\enquote {\bibinfo {title} {Dzyaloshinskii-moriya interaction at an
  antiferromagnetic interface: First-principles study of fe/ir bilayers on
  rh(001)},}\ }\href {\doibase 10.1103/PhysRevB.96.094408} {\bibfield
  {journal} {\bibinfo  {journal} {Phys. Rev. B}\ }\textbf {\bibinfo {volume}
  {96}},\ \bibinfo {pages} {094408} (\bibinfo {year} {2017})}\BibitemShut
  {NoStop}%
\bibitem [{\citenamefont {Haldar}\ \emph {et~al.}(2018)\citenamefont {Haldar},
  \citenamefont {von Malottki}, \citenamefont {Meyer}, \citenamefont
  {Bessarab},\ and\ \citenamefont {Heinze}}]{PhysRevB.98.060413}%
  \BibitemOpen
  \bibfield  {author} {\bibinfo {author} {\bibfnamefont {S}~\bibnamefont
  {Haldar}}, \bibinfo {author} {\bibfnamefont {S}~\bibnamefont {von Malottki}},
  \bibinfo {author} {\bibfnamefont {S}~\bibnamefont {Meyer}}, \bibinfo {author}
  {\bibfnamefont {P~F.}\ \bibnamefont {Bessarab}}, \ and\ \bibinfo {author}
  {\bibfnamefont {S}~\bibnamefont {Heinze}},\ }\bibfield  {title} {\enquote
  {\bibinfo {title} {First-principles prediction of sub-10-nm skyrmions in
  pd/fe bilayers on rh(111)},}\ }\href {\doibase 10.1103/PhysRevB.98.060413}
  {\bibfield  {journal} {\bibinfo  {journal} {Phys. Rev. B}\ }\textbf {\bibinfo
  {volume} {98}},\ \bibinfo {pages} {060413} (\bibinfo {year}
  {2018})}\BibitemShut {NoStop}%
\bibitem [{\citenamefont {Gutzeit}\ \emph {et~al.}(2021)\citenamefont
  {Gutzeit}, \citenamefont {Haldar}, \citenamefont {Meyer},\ and\ \citenamefont
  {Heinze}}]{PhysRevB.104.024420}%
  \BibitemOpen
  \bibfield  {author} {\bibinfo {author} {\bibfnamefont {M}~\bibnamefont
  {Gutzeit}}, \bibinfo {author} {\bibfnamefont {S}~\bibnamefont {Haldar}},
  \bibinfo {author} {\bibfnamefont {S}~\bibnamefont {Meyer}}, \ and\ \bibinfo
  {author} {\bibfnamefont {S}~\bibnamefont {Heinze}},\ }\bibfield  {title}
  {\enquote {\bibinfo {title} {Trends of higher-order exchange interactions in
  transition metal trilayers},}\ }\href {\doibase 10.1103/PhysRevB.104.024420}
  {\bibfield  {journal} {\bibinfo  {journal} {Phys. Rev. B}\ }\textbf {\bibinfo
  {volume} {104}},\ \bibinfo {pages} {024420} (\bibinfo {year}
  {2021})}\BibitemShut {NoStop}%
\bibitem [{\citenamefont {Nickel}\ \emph {et~al.}(2023)\citenamefont {Nickel},
  \citenamefont {Meyer},\ and\ \citenamefont {Heinze}}]{PhysRevB.107.174430}%
  \BibitemOpen
  \bibfield  {author} {\bibinfo {author} {\bibfnamefont {F}~\bibnamefont
  {Nickel}}, \bibinfo {author} {\bibfnamefont {S}~\bibnamefont {Meyer}}, \ and\
  \bibinfo {author} {\bibfnamefont {S}~\bibnamefont {Heinze}},\ }\bibfield
  {title} {\enquote {\bibinfo {title} {Exchange and dzyaloshinskii-moriya
  interaction in rh/co/fe/ir multilayers: Towards skyrmions in
  exchange-frustrated multilayers},}\ }\href {\doibase
  10.1103/PhysRevB.107.174430} {\bibfield  {journal} {\bibinfo  {journal}
  {Phys. Rev. B}\ }\textbf {\bibinfo {volume} {107}},\ \bibinfo {pages}
  {174430} (\bibinfo {year} {2023})}\BibitemShut {NoStop}%
\bibitem [{\citenamefont {Romming}\ \emph {et~al.}(2018)\citenamefont
  {Romming}, \citenamefont {Pralow}, \citenamefont {Kubetzka}, \citenamefont
  {Hoffmann}, \citenamefont {von M~Stephan}, \citenamefont {Meyer},
  \citenamefont {Dup\'e}, \citenamefont {Wiesendanger}, \citenamefont {von
  Bergmann},\ and\ \citenamefont {Heinze}}]{PhysRevLett.120.207201}%
  \BibitemOpen
  \bibfield  {author} {\bibinfo {author} {\bibfnamefont {N}~\bibnamefont
  {Romming}}, \bibinfo {author} {\bibfnamefont {He}~\bibnamefont {Pralow}},
  \bibinfo {author} {\bibfnamefont {A}~\bibnamefont {Kubetzka}}, \bibinfo
  {author} {\bibfnamefont {M}~\bibnamefont {Hoffmann}}, \bibinfo {author}
  {\bibnamefont {von M~Stephan}}, \bibinfo {author} {\bibfnamefont
  {S}~\bibnamefont {Meyer}}, \bibinfo {author} {\bibfnamefont {B}~\bibnamefont
  {Dup\'e}}, \bibinfo {author} {\bibfnamefont {R}~\bibnamefont {Wiesendanger}},
  \bibinfo {author} {\bibfnamefont {K}~\bibnamefont {von Bergmann}}, \ and\
  \bibinfo {author} {\bibfnamefont {S}~\bibnamefont {Heinze}},\ }\bibfield
  {title} {\enquote {\bibinfo {title} {Competition of dzyaloshinskii-moriya and
  higher-order exchange interactions in $\mathrm{Rh}/\mathrm{Fe}$ atomic
  bilayers on ir(111)},}\ }\href {\doibase 10.1103/PhysRevLett.120.207201}
  {\bibfield  {journal} {\bibinfo  {journal} {Phys. Rev. Lett.}\ }\textbf
  {\bibinfo {volume} {120}},\ \bibinfo {pages} {207201} (\bibinfo {year}
  {2018})}\BibitemShut {NoStop}%
\bibitem [{\citenamefont {Hasselberg}\ \emph {et~al.}(2015)\citenamefont
  {Hasselberg}, \citenamefont {Yanes}, \citenamefont {Hinzke}, \citenamefont
  {Sessi}, \citenamefont {Bode}, \citenamefont {Szunyogh},\ and\ \citenamefont
  {Nowak}}]{PhysRevB.91.064402}%
  \BibitemOpen
  \bibfield  {author} {\bibinfo {author} {\bibfnamefont {G.}~\bibnamefont
  {Hasselberg}}, \bibinfo {author} {\bibfnamefont {R.}~\bibnamefont {Yanes}},
  \bibinfo {author} {\bibfnamefont {D.}~\bibnamefont {Hinzke}}, \bibinfo
  {author} {\bibfnamefont {P.}~\bibnamefont {Sessi}}, \bibinfo {author}
  {\bibfnamefont {M.}~\bibnamefont {Bode}}, \bibinfo {author} {\bibfnamefont
  {L.}~\bibnamefont {Szunyogh}}, \ and\ \bibinfo {author} {\bibfnamefont
  {U.}~\bibnamefont {Nowak}},\ }\bibfield  {title} {\enquote {\bibinfo {title}
  {Thermal properties of a spin spiral: Manganese on tungsten(110)},}\ }\href
  {\doibase 10.1103/PhysRevB.91.064402} {\bibfield  {journal} {\bibinfo
  {journal} {Phys. Rev. B}\ }\textbf {\bibinfo {volume} {91}},\ \bibinfo
  {pages} {064402} (\bibinfo {year} {2015})}\BibitemShut {NoStop}%
\bibitem [{\citenamefont {Ferriani}\ \emph {et~al.}(2008)\citenamefont
  {Ferriani}, \citenamefont {von Bergmann}, \citenamefont {Vedmedenko},
  \citenamefont {Heinze}, \citenamefont {Bode}, \citenamefont {Heide},
  \citenamefont {Bihlmayer}, \citenamefont {Bl\"ugel},\ and\ \citenamefont
  {Wiesendanger}}]{PhysRevLett.101.027201}%
  \BibitemOpen
  \bibfield  {author} {\bibinfo {author} {\bibfnamefont {P.}~\bibnamefont
  {Ferriani}}, \bibinfo {author} {\bibfnamefont {K.}~\bibnamefont {von
  Bergmann}}, \bibinfo {author} {\bibfnamefont {E.~Y.}\ \bibnamefont
  {Vedmedenko}}, \bibinfo {author} {\bibfnamefont {S.}~\bibnamefont {Heinze}},
  \bibinfo {author} {\bibfnamefont {M.}~\bibnamefont {Bode}}, \bibinfo {author}
  {\bibfnamefont {M.}~\bibnamefont {Heide}}, \bibinfo {author} {\bibfnamefont
  {G.}~\bibnamefont {Bihlmayer}}, \bibinfo {author} {\bibfnamefont
  {S.}~\bibnamefont {Bl\"ugel}}, \ and\ \bibinfo {author} {\bibfnamefont
  {R.}~\bibnamefont {Wiesendanger}},\ }\bibfield  {title} {\enquote {\bibinfo
  {title} {Atomic-scale spin spiral with a unique rotational sense: Mn
  monolayer on w(001)},}\ }\href {\doibase 10.1103/PhysRevLett.101.027201}
  {\bibfield  {journal} {\bibinfo  {journal} {Phys. Rev. Lett.}\ }\textbf
  {\bibinfo {volume} {101}},\ \bibinfo {pages} {027201} (\bibinfo {year}
  {2008})}\BibitemShut {NoStop}%
\bibitem [{\citenamefont {Dup{\'e}}\ \emph {et~al.}(2014)\citenamefont
  {Dup{\'e}}, \citenamefont {Hoffmann}, \citenamefont {Paillard},\ and\
  \citenamefont {Heinze}}]{Dupe2014-ej}%
  \BibitemOpen
  \bibfield  {author} {\bibinfo {author} {\bibfnamefont {B}~\bibnamefont
  {Dup{\'e}}}, \bibinfo {author} {\bibfnamefont {M}~\bibnamefont {Hoffmann}},
  \bibinfo {author} {\bibfnamefont {C}~\bibnamefont {Paillard}}, \ and\
  \bibinfo {author} {\bibfnamefont {S}~\bibnamefont {Heinze}},\ }\bibfield
  {title} {\enquote {\bibinfo {title} {Tailoring magnetic skyrmions in
  ultra-thin transition metal films},}\ }\href
  {https://doi.org/10.1038/ncomms5030} {\bibfield  {journal} {\bibinfo
  {journal} {Nature Communications}\ }\textbf {\bibinfo {volume} {5}},\
  \bibinfo {pages} {4030} (\bibinfo {year} {2014})}\BibitemShut {NoStop}%
\bibitem [{\citenamefont {Simon}\ \emph {et~al.}(2014)\citenamefont {Simon},
  \citenamefont {Palot\'as}, \citenamefont {R\'ozsa}, \citenamefont {Udvardi},\
  and\ \citenamefont {Szunyogh}}]{PhysRevB.90.094410}%
  \BibitemOpen
  \bibfield  {author} {\bibinfo {author} {\bibfnamefont {E.}~\bibnamefont
  {Simon}}, \bibinfo {author} {\bibfnamefont {K.}~\bibnamefont {Palot\'as}},
  \bibinfo {author} {\bibfnamefont {L.}~\bibnamefont {R\'ozsa}}, \bibinfo
  {author} {\bibfnamefont {L.}~\bibnamefont {Udvardi}}, \ and\ \bibinfo
  {author} {\bibfnamefont {L.}~\bibnamefont {Szunyogh}},\ }\bibfield  {title}
  {\enquote {\bibinfo {title} {Formation of magnetic skyrmions with tunable
  properties in pdfe bilayer deposited on ir(111)},}\ }\href {\doibase
  10.1103/PhysRevB.90.094410} {\bibfield  {journal} {\bibinfo  {journal} {Phys.
  Rev. B}\ }\textbf {\bibinfo {volume} {90}},\ \bibinfo {pages} {094410}
  (\bibinfo {year} {2014})}\BibitemShut {NoStop}%
\bibitem [{\citenamefont {Sadhukhan}\ \emph {et~al.}(2023)\citenamefont
  {Sadhukhan}, \citenamefont {Bergman}, \citenamefont {Hellsvik}, \citenamefont
  {Thunstr{\"o}m},\ and\ \citenamefont {Delin}}]{sadhukhan2023spin}%
  \BibitemOpen
  \bibfield  {author} {\bibinfo {author} {\bibfnamefont {B}~\bibnamefont
  {Sadhukhan}}, \bibinfo {author} {\bibfnamefont {A}~\bibnamefont {Bergman}},
  \bibinfo {author} {\bibfnamefont {J}~\bibnamefont {Hellsvik}}, \bibinfo
  {author} {\bibfnamefont {P}~\bibnamefont {Thunstr{\"o}m}}, \ and\ \bibinfo
  {author} {\bibfnamefont {A}~\bibnamefont {Delin}},\ }\bibfield  {title}
  {\enquote {\bibinfo {title} {Spin-lattice couplings in a skyrmion multilayers
  of pd-fe/ir (111)},}\ }\href {https://doi.org/10.48550/arXiv.2309.03074}
  {\bibfield  {journal} {\bibinfo  {journal} {arXiv preprint arXiv:2309.03074}\
  } (\bibinfo {year} {2023})}\BibitemShut {NoStop}%
\bibitem [{\citenamefont {Miranda}\ \emph {et~al.}(2022)\citenamefont
  {Miranda}, \citenamefont {Klautau}, \citenamefont {Bergman},\ and\
  \citenamefont {Petrilli}}]{PhysRevB.105.224413}%
  \BibitemOpen
  \bibfield  {author} {\bibinfo {author} {\bibfnamefont {I.~P.}\ \bibnamefont
  {Miranda}}, \bibinfo {author} {\bibfnamefont {A.~B.}\ \bibnamefont
  {Klautau}}, \bibinfo {author} {\bibfnamefont {A.}~\bibnamefont {Bergman}}, \
  and\ \bibinfo {author} {\bibfnamefont {H.~M.}\ \bibnamefont {Petrilli}},\
  }\bibfield  {title} {\enquote {\bibinfo {title} {Band filling effects on the
  emergence of magnetic skyrmions: Pd/fe and pd/co bilayers on ir(111)},}\
  }\href {\doibase 10.1103/PhysRevB.105.224413} {\bibfield  {journal} {\bibinfo
   {journal} {Phys. Rev. B}\ }\textbf {\bibinfo {volume} {105}},\ \bibinfo
  {pages} {224413} (\bibinfo {year} {2022})}\BibitemShut {NoStop}%
\bibitem [{\citenamefont {von Malottki}\ \emph {et~al.}(2017)\citenamefont {von
  Malottki}, \citenamefont {Dup{\'e}}, \citenamefont {Bessarab}, \citenamefont
  {Delin},\ and\ \citenamefont {Heinze}}]{vonMalottki2017}%
  \BibitemOpen
  \bibfield  {author} {\bibinfo {author} {\bibfnamefont {S.}~\bibnamefont {von
  Malottki}}, \bibinfo {author} {\bibfnamefont {B.}~\bibnamefont {Dup{\'e}}},
  \bibinfo {author} {\bibfnamefont {P.~F.}\ \bibnamefont {Bessarab}}, \bibinfo
  {author} {\bibfnamefont {A.}~\bibnamefont {Delin}}, \ and\ \bibinfo {author}
  {\bibfnamefont {S.}~\bibnamefont {Heinze}},\ }\bibfield  {title} {\enquote
  {\bibinfo {title} {Enhanced skyrmion stability due to exchange
  frustration},}\ }\href {\doibase 10.1038/s41598-017-12525-x} {\bibfield
  {journal} {\bibinfo  {journal} {Scientific Reports}\ }\textbf {\bibinfo
  {volume} {7}},\ \bibinfo {pages} {12299} (\bibinfo {year}
  {2017})}\BibitemShut {NoStop}%
\bibitem [{\citenamefont {Moriya}(1960)}]{PhysRev.120.91}%
  \BibitemOpen
  \bibfield  {author} {\bibinfo {author} {\bibfnamefont {T}~\bibnamefont
  {Moriya}},\ }\bibfield  {title} {\enquote {\bibinfo {title} {Anisotropic
  superexchange interaction and weak ferromagnetism},}\ }\href {\doibase
  10.1103/PhysRev.120.91} {\bibfield  {journal} {\bibinfo  {journal} {Phys.
  Rev.}\ }\textbf {\bibinfo {volume} {120}},\ \bibinfo {pages} {91--98}
  (\bibinfo {year} {1960})}\BibitemShut {NoStop}%
\bibitem [{\citenamefont {Dzyaloshinsky}(1958)}]{DZYALOSHINSKY1958241}%
  \BibitemOpen
  \bibfield  {author} {\bibinfo {author} {\bibfnamefont {I.}~\bibnamefont
  {Dzyaloshinsky}},\ }\bibfield  {title} {\enquote {\bibinfo {title} {A
  thermodynamic theory of “weak” ferromagnetism of antiferromagnetics},}\
  }\href {\doibase https://doi.org/10.1016/0022-3697(58)90076-3} {\bibfield
  {journal} {\bibinfo  {journal} {Journal of Physics and Chemistry of Solids}\
  }\textbf {\bibinfo {volume} {4}},\ \bibinfo {pages} {241--255} (\bibinfo
  {year} {1958})}\BibitemShut {NoStop}%
\bibitem [{\citenamefont {Crépieux}\ and\ \citenamefont
  {Lacroix}(1998)}]{CREPIEUX1998341}%
  \BibitemOpen
  \bibfield  {author} {\bibinfo {author} {\bibfnamefont {A.}~\bibnamefont
  {Crépieux}}\ and\ \bibinfo {author} {\bibfnamefont {C.}~\bibnamefont
  {Lacroix}},\ }\bibfield  {title} {\enquote {\bibinfo {title}
  {Dzyaloshinsky–moriya interactions induced by symmetry breaking at a
  surface},}\ }\href {\doibase https://doi.org/10.1016/S0304-8853(97)01044-5}
  {\bibfield  {journal} {\bibinfo  {journal} {Journal of Magnetism and Magnetic
  Materials}\ }\textbf {\bibinfo {volume} {182}},\ \bibinfo {pages} {341--349}
  (\bibinfo {year} {1998})}\BibitemShut {NoStop}%
\bibitem [{\citenamefont {Sadhukhan}\ \emph {et~al.}(2024)\citenamefont
  {Sadhukhan}, \citenamefont {Bergman}, \citenamefont {Thunstr\"om},
  \citenamefont {L}, \citenamefont {Eriksson},\ and\ \citenamefont
  {Delin}}]{PhysRevB.110.174412}%
  \BibitemOpen
  \bibfield  {author} {\bibinfo {author} {\bibfnamefont {B}~\bibnamefont
  {Sadhukhan}}, \bibinfo {author} {\bibfnamefont {A}~\bibnamefont {Bergman}},
  \bibinfo {author} {\bibfnamefont {P}~\bibnamefont {Thunstr\"om}}, \bibinfo
  {author} {\bibfnamefont {M~Pereiro}\ \bibnamefont {L}}, \bibinfo {author}
  {\bibfnamefont {O}~\bibnamefont {Eriksson}}, \ and\ \bibinfo {author}
  {\bibfnamefont {A}~\bibnamefont {Delin}},\ }\bibfield  {title} {\enquote
  {\bibinfo {title} {Topological magnon in exchange frustration driven
  incommensurate spin spiral of kagome-lattice
  ${\mathrm{ymn}}_{6}{\mathrm{sn}}_{6}$},}\ }\href {\doibase
  10.1103/PhysRevB.110.174412} {\bibfield  {journal} {\bibinfo  {journal}
  {Phys. Rev. B}\ }\textbf {\bibinfo {volume} {110}},\ \bibinfo {pages}
  {174412} (\bibinfo {year} {2024})}\BibitemShut {NoStop}%
\bibitem [{\citenamefont {Bogdanov}\ and\ \citenamefont
  {Hubert}(1994)}]{BOGDANOV1994255}%
  \BibitemOpen
  \bibfield  {author} {\bibinfo {author} {\bibfnamefont {A.}~\bibnamefont
  {Bogdanov}}\ and\ \bibinfo {author} {\bibfnamefont {A.}~\bibnamefont
  {Hubert}},\ }\bibfield  {title} {\enquote {\bibinfo {title}
  {Thermodynamically stable magnetic vortex states in magnetic crystals},}\
  }\href {\doibase https://doi.org/10.1016/0304-8853(94)90046-9} {\bibfield
  {journal} {\bibinfo  {journal} {Journal of Magnetism and Magnetic Materials}\
  }\textbf {\bibinfo {volume} {138}},\ \bibinfo {pages} {255--269} (\bibinfo
  {year} {1994})}\BibitemShut {NoStop}%
\bibitem [{\citenamefont {R{\"o}{\ss}ler}\ \emph {et~al.}(2006)\citenamefont
  {R{\"o}{\ss}ler}, \citenamefont {Bogdanov},\ and\ \citenamefont
  {Pfleiderer}}]{Rosler2006-jf}%
  \BibitemOpen
  \bibfield  {author} {\bibinfo {author} {\bibfnamefont {U~K}\ \bibnamefont
  {R{\"o}{\ss}ler}}, \bibinfo {author} {\bibfnamefont {A~N}\ \bibnamefont
  {Bogdanov}}, \ and\ \bibinfo {author} {\bibfnamefont {C}~\bibnamefont
  {Pfleiderer}},\ }\bibfield  {title} {\enquote {\bibinfo {title} {Spontaneous
  skyrmion ground states in magnetic metals},}\ }\href
  {https://doi.org/10.1038/nature05056} {\bibfield  {journal} {\bibinfo
  {journal} {Nature}\ }\textbf {\bibinfo {volume} {442}},\ \bibinfo {pages}
  {797--801} (\bibinfo {year} {2006})}\BibitemShut {NoStop}%
\bibitem [{\citenamefont {Dup{\'e}}\ \emph {et~al.}(2016)\citenamefont
  {Dup{\'e}}, \citenamefont {Bihlmayer}, \citenamefont {B{\"o}ttcher},
  \citenamefont {Bl{\"u}gel},\ and\ \citenamefont {Heinze}}]{Dupe2016-wp}%
  \BibitemOpen
  \bibfield  {author} {\bibinfo {author} {\bibfnamefont {B}~\bibnamefont
  {Dup{\'e}}}, \bibinfo {author} {\bibfnamefont {G}~\bibnamefont {Bihlmayer}},
  \bibinfo {author} {\bibfnamefont {M}~\bibnamefont {B{\"o}ttcher}}, \bibinfo
  {author} {\bibfnamefont {S}~\bibnamefont {Bl{\"u}gel}}, \ and\ \bibinfo
  {author} {\bibfnamefont {S}~\bibnamefont {Heinze}},\ }\bibfield  {title}
  {\enquote {\bibinfo {title} {Engineering skyrmions in transition-metal
  multilayers for spintronics},}\ }\href {https://doi.org/10.1038/ncomms11779}
  {\bibfield  {journal} {\bibinfo  {journal} {Nature Communications}\ }\textbf
  {\bibinfo {volume} {7}},\ \bibinfo {pages} {11779} (\bibinfo {year}
  {2016})}\BibitemShut {NoStop}%
\bibitem [{\citenamefont {R\'ozsa}\ \emph
  {et~al.}(2016{\natexlab{a}})\citenamefont {R\'ozsa}, \citenamefont {De\'ak},
  \citenamefont {Simon}, \citenamefont {Yanes}, \citenamefont {Udvardi},
  \citenamefont {Szunyogh},\ and\ \citenamefont
  {Nowak}}]{PhysRevLett.117.157205}%
  \BibitemOpen
  \bibfield  {author} {\bibinfo {author} {\bibfnamefont {L}~\bibnamefont
  {R\'ozsa}}, \bibinfo {author} {\bibfnamefont {A}~\bibnamefont {De\'ak}},
  \bibinfo {author} {\bibfnamefont {E}~\bibnamefont {Simon}}, \bibinfo {author}
  {\bibfnamefont {R}~\bibnamefont {Yanes}}, \bibinfo {author} {\bibfnamefont
  {L}~\bibnamefont {Udvardi}}, \bibinfo {author} {\bibfnamefont
  {L}~\bibnamefont {Szunyogh}}, \ and\ \bibinfo {author} {\bibfnamefont
  {U}~\bibnamefont {Nowak}},\ }\bibfield  {title} {\enquote {\bibinfo {title}
  {Skyrmions with attractive interactions in an ultrathin magnetic film},}\
  }\href {\doibase 10.1103/PhysRevLett.117.157205} {\bibfield  {journal}
  {\bibinfo  {journal} {Phys. Rev. Lett.}\ }\textbf {\bibinfo {volume} {117}},\
  \bibinfo {pages} {157205} (\bibinfo {year} {2016}{\natexlab{a}})}\BibitemShut
  {NoStop}%
\bibitem [{\citenamefont {Hrabec}\ \emph {et~al.}(2014)\citenamefont {Hrabec},
  \citenamefont {Porter}, \citenamefont {Wells}, \citenamefont {Benitez},
  \citenamefont {Burnell}, \citenamefont {McVitie}, \citenamefont {McGrouther},
  \citenamefont {Moore},\ and\ \citenamefont {Marrows}}]{PhysRevB.90.020402}%
  \BibitemOpen
  \bibfield  {author} {\bibinfo {author} {\bibfnamefont {A.}~\bibnamefont
  {Hrabec}}, \bibinfo {author} {\bibfnamefont {N.~A.}\ \bibnamefont {Porter}},
  \bibinfo {author} {\bibfnamefont {A.}~\bibnamefont {Wells}}, \bibinfo
  {author} {\bibfnamefont {M.~J.}\ \bibnamefont {Benitez}}, \bibinfo {author}
  {\bibfnamefont {G.}~\bibnamefont {Burnell}}, \bibinfo {author} {\bibfnamefont
  {S.}~\bibnamefont {McVitie}}, \bibinfo {author} {\bibfnamefont
  {D.}~\bibnamefont {McGrouther}}, \bibinfo {author} {\bibfnamefont {T.~A.}\
  \bibnamefont {Moore}}, \ and\ \bibinfo {author} {\bibfnamefont {C.~H.}\
  \bibnamefont {Marrows}},\ }\bibfield  {title} {\enquote {\bibinfo {title}
  {Measuring and tailoring the dzyaloshinskii-moriya interaction in
  perpendicularly magnetized thin films},}\ }\href {\doibase
  10.1103/PhysRevB.90.020402} {\bibfield  {journal} {\bibinfo  {journal} {Phys.
  Rev. B}\ }\textbf {\bibinfo {volume} {90}},\ \bibinfo {pages} {020402}
  (\bibinfo {year} {2014})}\BibitemShut {NoStop}%
\bibitem [{\citenamefont {Hardrat}\ \emph {et~al.}(2009)\citenamefont
  {Hardrat}, \citenamefont {Al-Zubi}, \citenamefont {Ferriani}, \citenamefont
  {Bl\"ugel}, \citenamefont {Bihlmayer},\ and\ \citenamefont
  {Heinze}}]{PhysRevB.79.094411}%
  \BibitemOpen
  \bibfield  {author} {\bibinfo {author} {\bibfnamefont {B.}~\bibnamefont
  {Hardrat}}, \bibinfo {author} {\bibfnamefont {A.}~\bibnamefont {Al-Zubi}},
  \bibinfo {author} {\bibfnamefont {P.}~\bibnamefont {Ferriani}}, \bibinfo
  {author} {\bibfnamefont {S.}~\bibnamefont {Bl\"ugel}}, \bibinfo {author}
  {\bibfnamefont {G.}~\bibnamefont {Bihlmayer}}, \ and\ \bibinfo {author}
  {\bibfnamefont {S.}~\bibnamefont {Heinze}},\ }\bibfield  {title} {\enquote
  {\bibinfo {title} {Complex magnetism of iron monolayers on hexagonal
  transition metal surfaces from first principles},}\ }\href {\doibase
  10.1103/PhysRevB.79.094411} {\bibfield  {journal} {\bibinfo  {journal} {Phys.
  Rev. B}\ }\textbf {\bibinfo {volume} {79}},\ \bibinfo {pages} {094411}
  (\bibinfo {year} {2009})}\BibitemShut {NoStop}%
\bibitem [{\citenamefont {Herv{\'e}}\ \emph {et~al.}(2018)\citenamefont
  {Herv{\'e}}, \citenamefont {Dup{\'e}}, \citenamefont {Lopes}, \citenamefont
  {B{\"o}ttcher}, \citenamefont {Martins}, \citenamefont {Balashov},
  \citenamefont {Gerhard}, \citenamefont {Sinova},\ and\ \citenamefont
  {Wulfhekel}}]{Herve2018-qo}%
  \BibitemOpen
  \bibfield  {author} {\bibinfo {author} {\bibfnamefont {M}~\bibnamefont
  {Herv{\'e}}}, \bibinfo {author} {\bibfnamefont {B}~\bibnamefont {Dup{\'e}}},
  \bibinfo {author} {\bibfnamefont {R}~\bibnamefont {Lopes}}, \bibinfo {author}
  {\bibfnamefont {M}~\bibnamefont {B{\"o}ttcher}}, \bibinfo {author}
  {\bibfnamefont {M~D}\ \bibnamefont {Martins}}, \bibinfo {author}
  {\bibfnamefont {T}~\bibnamefont {Balashov}}, \bibinfo {author} {\bibfnamefont
  {L}~\bibnamefont {Gerhard}}, \bibinfo {author} {\bibfnamefont
  {J}~\bibnamefont {Sinova}}, \ and\ \bibinfo {author} {\bibfnamefont
  {W}~\bibnamefont {Wulfhekel}},\ }\bibfield  {title} {\enquote {\bibinfo
  {title} {Stabilizing spin spirals and isolated skyrmions at low magnetic
  field exploiting vanishing magnetic anisotropy},}\ }\href
  {https://doi.org/10.1038/s41467-018-03240-w} {\bibfield  {journal} {\bibinfo
  {journal} {Nature Communications}\ }\textbf {\bibinfo {volume} {9}},\
  \bibinfo {pages} {1015} (\bibinfo {year} {2018})}\BibitemShut {NoStop}%
\bibitem [{\citenamefont {Meyer}\ \emph {et~al.}(2019)\citenamefont {Meyer},
  \citenamefont {Perini}, \citenamefont {von Malottki}, \citenamefont
  {Kubetzka}, \citenamefont {Wiesendanger}, \citenamefont {von Bergmann},\ and\
  \citenamefont {Heinze}}]{Meyer2019-ui}%
  \BibitemOpen
  \bibfield  {author} {\bibinfo {author} {\bibfnamefont {S}~\bibnamefont
  {Meyer}}, \bibinfo {author} {\bibfnamefont {M}~\bibnamefont {Perini}},
  \bibinfo {author} {\bibfnamefont {S}~\bibnamefont {von Malottki}}, \bibinfo
  {author} {\bibfnamefont {A}~\bibnamefont {Kubetzka}}, \bibinfo {author}
  {\bibfnamefont {R}~\bibnamefont {Wiesendanger}}, \bibinfo {author}
  {\bibfnamefont {K}~\bibnamefont {von Bergmann}}, \ and\ \bibinfo {author}
  {\bibfnamefont {S}~\bibnamefont {Heinze}},\ }\bibfield  {title} {\enquote
  {\bibinfo {title} {Isolated zero field sub-10 nm skyrmions in ultrathin co
  films},}\ }\href {https://doi.org/10.1038/s41467-019-11831-4} {\bibfield
  {journal} {\bibinfo  {journal} {Nature Communications}\ }\textbf {\bibinfo
  {volume} {10}},\ \bibinfo {pages} {3823} (\bibinfo {year}
  {2019})}\BibitemShut {NoStop}%
\bibitem [{\citenamefont {Kresse}\ and\ \citenamefont
  {Joubert}(1999)}]{PhysRevB.59.1758}%
  \BibitemOpen
  \bibfield  {author} {\bibinfo {author} {\bibfnamefont {G.}~\bibnamefont
  {Kresse}}\ and\ \bibinfo {author} {\bibfnamefont {D.}~\bibnamefont
  {Joubert}},\ }\bibfield  {title} {\enquote {\bibinfo {title} {From ultrasoft
  pseudopotentials to the projector augmented-wave method},}\ }\href {\doibase
  10.1103/PhysRevB.59.1758} {\bibfield  {journal} {\bibinfo  {journal} {Phys.
  Rev. B}\ }\textbf {\bibinfo {volume} {59}},\ \bibinfo {pages} {1758--1775}
  (\bibinfo {year} {1999})}\BibitemShut {NoStop}%
\bibitem [{\citenamefont {Kresse}\ and\ \citenamefont
  {Furthm\"uller}(1996)}]{PhysRevB.54.11169}%
  \BibitemOpen
  \bibfield  {author} {\bibinfo {author} {\bibfnamefont {G.}~\bibnamefont
  {Kresse}}\ and\ \bibinfo {author} {\bibfnamefont {J.}~\bibnamefont
  {Furthm\"uller}},\ }\bibfield  {title} {\enquote {\bibinfo {title} {Efficient
  iterative schemes for ab initio total-energy calculations using a plane-wave
  basis set},}\ }\href {\doibase 10.1103/PhysRevB.54.11169} {\bibfield
  {journal} {\bibinfo  {journal} {Phys. Rev. B}\ }\textbf {\bibinfo {volume}
  {54}},\ \bibinfo {pages} {11169--11186} (\bibinfo {year} {1996})}\BibitemShut
  {NoStop}%
\bibitem [{vas()}]{vasp}%
  \BibitemOpen
  \href@noop {} {}\bibinfo {note} {Https://www.vasp.at}\BibitemShut {NoStop}%
\bibitem [{\citenamefont {energy}\ \emph {et~al.}(2010)\citenamefont {energy},
  \citenamefont {force calculations with~density functional}, \citenamefont
  {dynamical mean~field theory}, \citenamefont {Alouani}, \citenamefont
  {Andersson}, \citenamefont {Delin}, \citenamefont {Eriksson},\ and\
  \citenamefont {Grechnyev}}]{wills2010full}%
  \BibitemOpen
  \bibfield  {author} {\bibinfo {author} {\bibfnamefont {Full-Potential
  Electronic Structure~Method:}\ \bibnamefont {energy}}, \bibinfo {author}
  {\bibnamefont {force calculations with~density functional}}, \bibinfo
  {author} {\bibfnamefont {John~M}\ \bibnamefont {dynamical mean~field theory},
  \bibfnamefont {Wills}}, \bibinfo {author} {\bibfnamefont {M}~\bibnamefont
  {Alouani}}, \bibinfo {author} {\bibfnamefont {P}~\bibnamefont {Andersson}},
  \bibinfo {author} {\bibfnamefont {A}~\bibnamefont {Delin}}, \bibinfo {author}
  {\bibfnamefont {O}~\bibnamefont {Eriksson}}, \ and\ \bibinfo {author}
  {\bibfnamefont {O}~\bibnamefont {Grechnyev}},\ }\href {\doibase
  10.1007/978-3-642-15144-6} {}Vol.\ \bibinfo {volume} {167}\ (\bibinfo
  {publisher} {Springer Science \& Business Media},\ \bibinfo {year}
  {2010})\BibitemShut {NoStop}%
\bibitem [{\citenamefont {relativistic spin~polarized toolkit}()}]{rsptweb}%
  \BibitemOpen
  \bibfield  {author} {\bibinfo {author} {\bibfnamefont {RSPt}\ \bibnamefont
  {relativistic spin~polarized toolkit}},\ }\href {http://fplmto-rspt.org/}
  {\enquote {\bibinfo {title} {http://fplmto-rspt.org/},}\ }\BibitemShut
  {NoStop}%
\bibitem [{\citenamefont {Eriksson}\ \emph {et~al.}(2017)\citenamefont
  {Eriksson}, \citenamefont {Bergman}, \citenamefont {Bergqvist},\ and\
  \citenamefont {Hellsvik}}]{UppASD_book}%
  \BibitemOpen
  \bibfield  {author} {\bibinfo {author} {\bibfnamefont {O}~\bibnamefont
  {Eriksson}}, \bibinfo {author} {\bibfnamefont {A}~\bibnamefont {Bergman}},
  \bibinfo {author} {\bibfnamefont {L}~\bibnamefont {Bergqvist}}, \ and\
  \bibinfo {author} {\bibfnamefont {J}~\bibnamefont {Hellsvik}},\ }\href@noop
  {} {\emph {\bibinfo {title} {Atomistic spin dynamics: Foundations and
  applications}}}\ (\bibinfo  {publisher} {Oxford university press},\ \bibinfo
  {year} {2017})\BibitemShut {NoStop}%
\bibitem [{upp()}]{uppasd}%
  \BibitemOpen
  \href@noop {} {}\bibinfo {note} {Http://github.com/UppASD/UppASD}\BibitemShut
  {NoStop}%
\bibitem [{\citenamefont {Sadhukhan}\ \emph {et~al.}(2022)\citenamefont
  {Sadhukhan}, \citenamefont {Bergman}, \citenamefont {Kvashnin}, \citenamefont
  {Hellsvik},\ and\ \citenamefont {Delin}}]{PhysRevB.105.104418}%
  \BibitemOpen
  \bibfield  {author} {\bibinfo {author} {\bibfnamefont {B}~\bibnamefont
  {Sadhukhan}}, \bibinfo {author} {\bibfnamefont {A}~\bibnamefont {Bergman}},
  \bibinfo {author} {\bibfnamefont {Y~O.}\ \bibnamefont {Kvashnin}}, \bibinfo
  {author} {\bibfnamefont {J}~\bibnamefont {Hellsvik}}, \ and\ \bibinfo
  {author} {\bibfnamefont {A}~\bibnamefont {Delin}},\ }\bibfield  {title}
  {\enquote {\bibinfo {title} {Spin-lattice couplings in two-dimensional
  ${\mathrm{cri}}_{3}$ from first-principles computations},}\ }\href {\doibase
  10.1103/PhysRevB.105.104418} {\bibfield  {journal} {\bibinfo  {journal}
  {Phys. Rev. B}\ }\textbf {\bibinfo {volume} {105}},\ \bibinfo {pages}
  {104418} (\bibinfo {year} {2022})}\BibitemShut {NoStop}%
\bibitem [{\citenamefont {Olive}\ \emph {et~al.}(1986)\citenamefont {Olive},
  \citenamefont {Young},\ and\ \citenamefont {Sherrington}}]{PhysRevB.34.6341}%
  \BibitemOpen
  \bibfield  {author} {\bibinfo {author} {\bibfnamefont {J.~A.}\ \bibnamefont
  {Olive}}, \bibinfo {author} {\bibfnamefont {A.~P.}\ \bibnamefont {Young}}, \
  and\ \bibinfo {author} {\bibfnamefont {D.}~\bibnamefont {Sherrington}},\
  }\bibfield  {title} {\enquote {\bibinfo {title} {Computer simulation of the
  three-dimensional short-range heisenberg spin glass},}\ }\href {\doibase
  10.1103/PhysRevB.34.6341} {\bibfield  {journal} {\bibinfo  {journal} {Phys.
  Rev. B}\ }\textbf {\bibinfo {volume} {34}},\ \bibinfo {pages} {6341--6346}
  (\bibinfo {year} {1986})}\BibitemShut {NoStop}%
\bibitem [{\citenamefont {Finco}\ \emph {et~al.}(2017)\citenamefont {Finco},
  \citenamefont {R\'ozsa}, \citenamefont {Hsu}, \citenamefont {Kubetzka},
  \citenamefont {Vedmedenko}, \citenamefont {von Bergmann},\ and\ \citenamefont
  {Wiesendanger}}]{PhysRevLett.119.037202}%
  \BibitemOpen
  \bibfield  {author} {\bibinfo {author} {\bibfnamefont {A}~\bibnamefont
  {Finco}}, \bibinfo {author} {\bibfnamefont {L}~\bibnamefont {R\'ozsa}},
  \bibinfo {author} {\bibfnamefont {P-J}\ \bibnamefont {Hsu}}, \bibinfo
  {author} {\bibfnamefont {A}~\bibnamefont {Kubetzka}}, \bibinfo {author}
  {\bibfnamefont {E}~\bibnamefont {Vedmedenko}}, \bibinfo {author}
  {\bibfnamefont {K}~\bibnamefont {von Bergmann}}, \ and\ \bibinfo {author}
  {\bibfnamefont {R}~\bibnamefont {Wiesendanger}},\ }\bibfield  {title}
  {\enquote {\bibinfo {title} {Temperature-induced increase of spin spiral
  periods},}\ }\href {\doibase 10.1103/PhysRevLett.119.037202} {\bibfield
  {journal} {\bibinfo  {journal} {Phys. Rev. Lett.}\ }\textbf {\bibinfo
  {volume} {119}},\ \bibinfo {pages} {037202} (\bibinfo {year}
  {2017})}\BibitemShut {NoStop}%
\bibitem [{\citenamefont {Okubo}\ \emph {et~al.}(2012)\citenamefont {Okubo},
  \citenamefont {Chung},\ and\ \citenamefont
  {Kawamura}}]{PhysRevLett.108.017206}%
  \BibitemOpen
  \bibfield  {author} {\bibinfo {author} {\bibfnamefont {T}~\bibnamefont
  {Okubo}}, \bibinfo {author} {\bibfnamefont {S}~\bibnamefont {Chung}}, \ and\
  \bibinfo {author} {\bibfnamefont {H}~\bibnamefont {Kawamura}},\ }\bibfield
  {title} {\enquote {\bibinfo {title} {Multiple-$q$ states and the skyrmion
  lattice of the triangular-lattice heisenberg antiferromagnet under magnetic
  fields},}\ }\href {\doibase 10.1103/PhysRevLett.108.017206} {\bibfield
  {journal} {\bibinfo  {journal} {Phys. Rev. Lett.}\ }\textbf {\bibinfo
  {volume} {108}},\ \bibinfo {pages} {017206} (\bibinfo {year}
  {2012})}\BibitemShut {NoStop}%
\bibitem [{\citenamefont {Buhrandt}\ and\ \citenamefont
  {Fritz}(2013)}]{PhysRevB.88.195137}%
  \BibitemOpen
  \bibfield  {author} {\bibinfo {author} {\bibfnamefont {S}~\bibnamefont
  {Buhrandt}}\ and\ \bibinfo {author} {\bibfnamefont {L}~\bibnamefont
  {Fritz}},\ }\bibfield  {title} {\enquote {\bibinfo {title} {Skyrmion lattice
  phase in three-dimensional chiral magnets from monte carlo simulations},}\
  }\href {\doibase 10.1103/PhysRevB.88.195137} {\bibfield  {journal} {\bibinfo
  {journal} {Phys. Rev. B}\ }\textbf {\bibinfo {volume} {88}},\ \bibinfo
  {pages} {195137} (\bibinfo {year} {2013})}\BibitemShut {NoStop}%
\bibitem [{\citenamefont {R\'ozsa}\ \emph
  {et~al.}(2016{\natexlab{b}})\citenamefont {R\'ozsa}, \citenamefont {Simon},
  \citenamefont {Palot\'as}, \citenamefont {Udvardi},\ and\ \citenamefont
  {Szunyogh}}]{PhysRevB.93.024417}%
  \BibitemOpen
  \bibfield  {author} {\bibinfo {author} {\bibfnamefont {L}~\bibnamefont
  {R\'ozsa}}, \bibinfo {author} {\bibfnamefont {E}~\bibnamefont {Simon}},
  \bibinfo {author} {\bibfnamefont {K}~\bibnamefont {Palot\'as}}, \bibinfo
  {author} {\bibfnamefont {L}~\bibnamefont {Udvardi}}, \ and\ \bibinfo {author}
  {\bibfnamefont {L}~\bibnamefont {Szunyogh}},\ }\bibfield  {title} {\enquote
  {\bibinfo {title} {Complex magnetic phase diagram and skyrmion lifetime in an
  ultrathin film from atomistic simulations},}\ }\href {\doibase
  10.1103/PhysRevB.93.024417} {\bibfield  {journal} {\bibinfo  {journal} {Phys.
  Rev. B}\ }\textbf {\bibinfo {volume} {93}},\ \bibinfo {pages} {024417}
  (\bibinfo {year} {2016}{\natexlab{b}})}\BibitemShut {NoStop}%
\bibitem [{\citenamefont {Böttcher}\ \emph {et~al.}(2018)\citenamefont
  {Böttcher}, \citenamefont {Heinze}, \citenamefont {Egorov}, \citenamefont
  {Sinova},\ and\ \citenamefont {Dupé}}]{Bottcher_2018}%
  \BibitemOpen
  \bibfield  {author} {\bibinfo {author} {\bibfnamefont {M}~\bibnamefont
  {Böttcher}}, \bibinfo {author} {\bibfnamefont {S}~\bibnamefont {Heinze}},
  \bibinfo {author} {\bibfnamefont {S}~\bibnamefont {Egorov}}, \bibinfo
  {author} {\bibfnamefont {J}~\bibnamefont {Sinova}}, \ and\ \bibinfo {author}
  {\bibfnamefont {B}~\bibnamefont {Dupé}},\ }\bibfield  {title} {\enquote
  {\bibinfo {title} {B–t phase diagram of pd/fe/ir(111) computed with
  parallel tempering monte carlo},}\ }\href {\doibase 10.1088/1367-2630/aae282}
  {\bibfield  {journal} {\bibinfo  {journal} {New Journal of Physics}\ }\textbf
  {\bibinfo {volume} {20}},\ \bibinfo {pages} {103014} (\bibinfo {year}
  {2018})}\BibitemShut {NoStop}%
\bibitem [{\citenamefont {Romming}\ \emph {et~al.}(2015)\citenamefont
  {Romming}, \citenamefont {Kubetzka}, \citenamefont {Hanneken}, \citenamefont
  {von Bergmann},\ and\ \citenamefont {Wiesendanger}}]{PhysRevLett.114.177203}%
  \BibitemOpen
  \bibfield  {author} {\bibinfo {author} {\bibfnamefont {N}~\bibnamefont
  {Romming}}, \bibinfo {author} {\bibfnamefont {A}~\bibnamefont {Kubetzka}},
  \bibinfo {author} {\bibfnamefont {C}~\bibnamefont {Hanneken}}, \bibinfo
  {author} {\bibfnamefont {K}~\bibnamefont {von Bergmann}}, \ and\ \bibinfo
  {author} {\bibfnamefont {R}~\bibnamefont {Wiesendanger}},\ }\bibfield
  {title} {\enquote {\bibinfo {title} {Field-dependent size and shape of single
  magnetic skyrmions},}\ }\href {\doibase 10.1103/PhysRevLett.114.177203}
  {\bibfield  {journal} {\bibinfo  {journal} {Phys. Rev. Lett.}\ }\textbf
  {\bibinfo {volume} {114}},\ \bibinfo {pages} {177203} (\bibinfo {year}
  {2015})}\BibitemShut {NoStop}%
\bibitem [{\citenamefont {Kwon}\ \emph {et~al.}(2020)\citenamefont {Kwon},
  \citenamefont {Song}, \citenamefont {Jeong}, \citenamefont {Lee},
  \citenamefont {Park}, \citenamefont {Kim}, \citenamefont {Won}, \citenamefont
  {Min}, \citenamefont {Chang},\ and\ \citenamefont {Choi}}]{Kwon2020}%
  \BibitemOpen
  \bibfield  {author} {\bibinfo {author} {\bibfnamefont {Hee~Young}\
  \bibnamefont {Kwon}}, \bibinfo {author} {\bibfnamefont {Kyung~Mee}\
  \bibnamefont {Song}}, \bibinfo {author} {\bibfnamefont {Juyoung}\
  \bibnamefont {Jeong}}, \bibinfo {author} {\bibfnamefont {Ah-Yeon}\
  \bibnamefont {Lee}}, \bibinfo {author} {\bibfnamefont {Seung-Young}\
  \bibnamefont {Park}}, \bibinfo {author} {\bibfnamefont {Jeehoon}\
  \bibnamefont {Kim}}, \bibinfo {author} {\bibfnamefont {Changyeon}\
  \bibnamefont {Won}}, \bibinfo {author} {\bibfnamefont {Byoung-Chul}\
  \bibnamefont {Min}}, \bibinfo {author} {\bibfnamefont {Hye~Jung}\
  \bibnamefont {Chang}}, \ and\ \bibinfo {author} {\bibfnamefont {Jun~Woo}\
  \bibnamefont {Choi}},\ }\bibfield  {title} {\enquote {\bibinfo {title}
  {High-density n{\'e}el-type magnetic skyrmion phase stabilized at high
  temperature},}\ }\href {\doibase 10.1038/s41427-020-00270-z} {\bibfield
  {journal} {\bibinfo  {journal} {NPG Asia Materials}\ }\textbf {\bibinfo
  {volume} {12}},\ \bibinfo {pages} {86} (\bibinfo {year} {2020})}\BibitemShut
  {NoStop}%
\bibitem [{\citenamefont {Sonntag}\ \emph {et~al.}(2014)\citenamefont
  {Sonntag}, \citenamefont {Hermenau}, \citenamefont {Krause},\ and\
  \citenamefont {Wiesendanger}}]{PhysRevLett.113.077202}%
  \BibitemOpen
  \bibfield  {author} {\bibinfo {author} {\bibfnamefont {A.}~\bibnamefont
  {Sonntag}}, \bibinfo {author} {\bibfnamefont {J.}~\bibnamefont {Hermenau}},
  \bibinfo {author} {\bibfnamefont {S.}~\bibnamefont {Krause}}, \ and\ \bibinfo
  {author} {\bibfnamefont {R.}~\bibnamefont {Wiesendanger}},\ }\bibfield
  {title} {\enquote {\bibinfo {title} {Thermal stability of an
  interface-stabilized skyrmion lattice},}\ }\href {\doibase
  10.1103/PhysRevLett.113.077202} {\bibfield  {journal} {\bibinfo  {journal}
  {Phys. Rev. Lett.}\ }\textbf {\bibinfo {volume} {113}},\ \bibinfo {pages}
  {077202} (\bibinfo {year} {2014})}\BibitemShut {NoStop}%
\end{thebibliography}%

\section{Appendix}

\begin{figure*} [t!] 
\centering
\includegraphics[width=1.03\textwidth,angle=0]{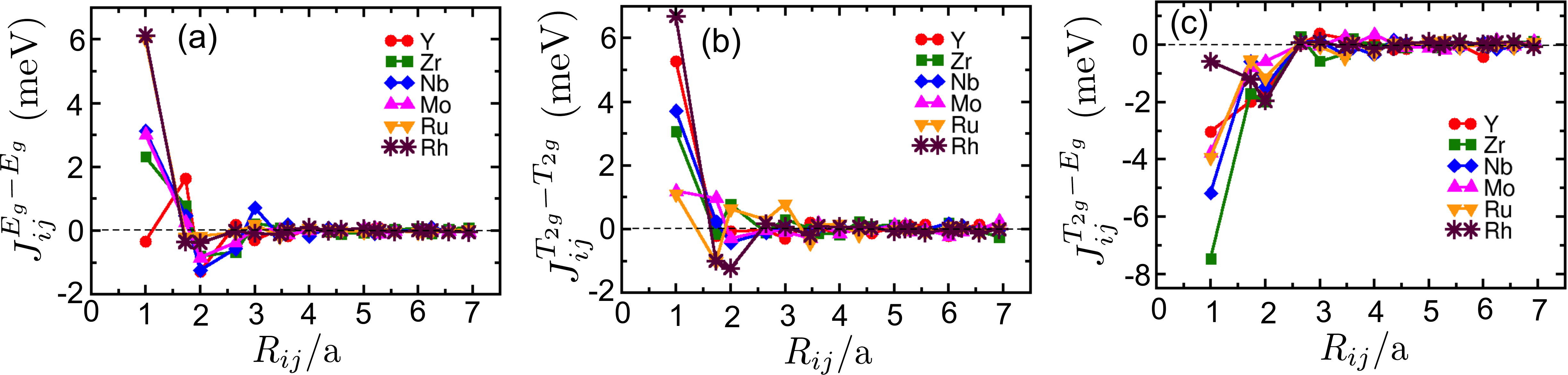} 
\caption{Orbital decomposition of magnetic exchange interactions of Fe-3d (a) ${J_{ij}}^{e_{g}-e_{g}}$,  (b) ${J_{ij}}^{t_{2g}-t_{2g}}$ and (c) ${J_{ij}}^{t_{2g}-e_{g}}$ in 4d/Fe/Ir(111) multilayers with 4d = Y, Zr, Nb, Mo, Ru and Rh.}
\label{sfig1} 
\end{figure*}

\begin{figure} [ht] 
\centering
\includegraphics[width=0.4\textwidth,angle=0]{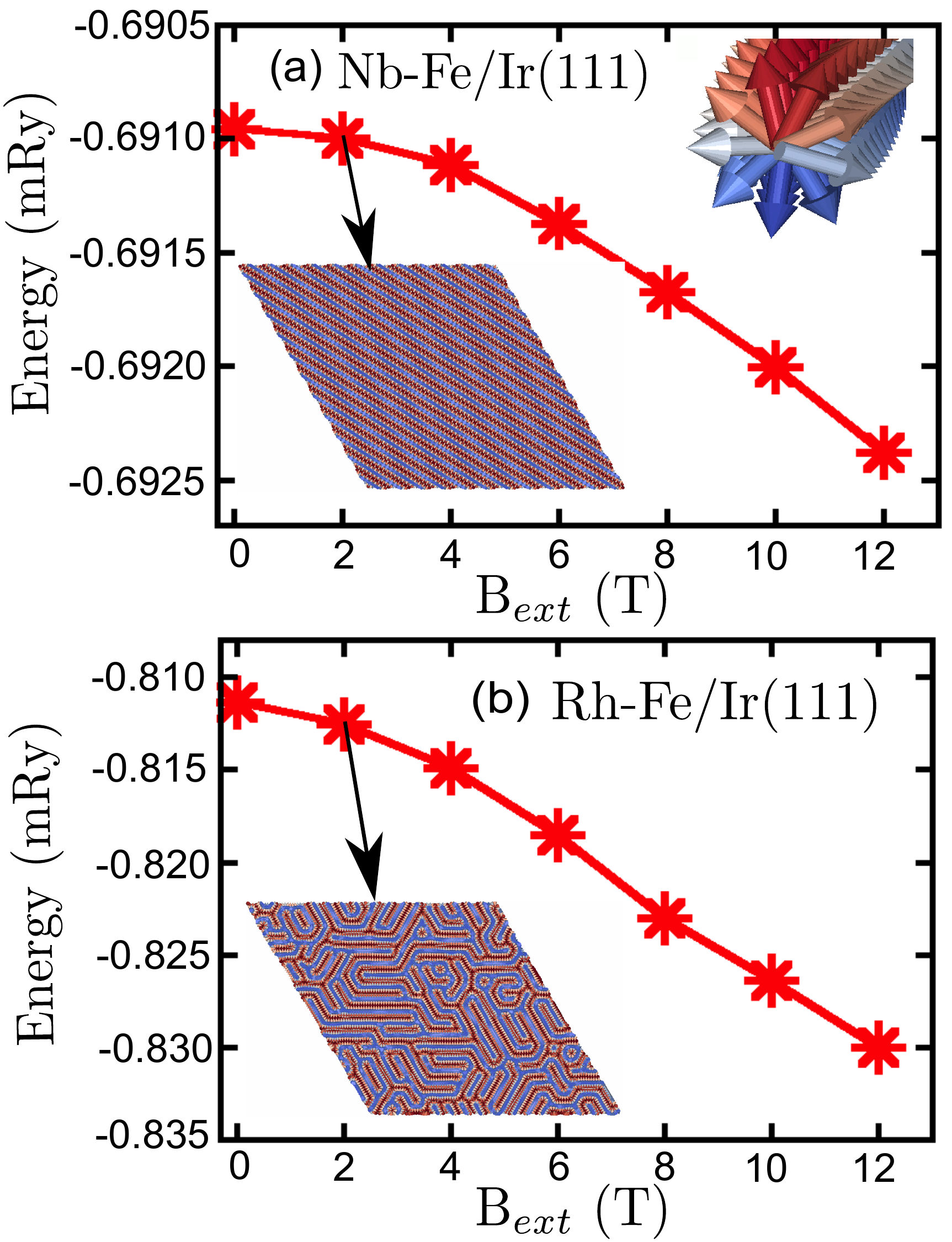} 
\caption{Simulated energies (energy/spin compared to its FM phase {\bl{(i.e E$_{\mathrm{MC}}$-E$_{\mathrm{FM}}$)}}) and spin textures of (a) Nb/Fe/Ir(111) and (b) Rh/Fe/Ir(111) with external magnetic fields from Monte Carlo simulations. }
\label{sfig2} 
\end{figure}

\par We investigate the orbital decomposition of isotropic Heisenberg exchange interactions to study the interfacial hybridization effects in 4d/Fe/Ir(111) multilayers.  {\bs{ In the case of 4d/Fe/Ir(111),  the Fe atoms are surrounded by Pd and Ir atoms.  The d-orbitals (Fe-3d,  X-4d and Ir-5d) are mostly playing the role in magnetic interactions of transition metal multilayers of 4d/Fe/Ir(111).  For simplicity,  we assign the Fe-3d orbitals to t$_{2g}$ and e$_g$ subsets which would arise in the ideal case.  The Fe atoms reside in an octahedral environment (C$_{3v}$ symmetry) in  4d/Fe/Ir(111) multilayers.  The crystal field interaction with the Pd and Ir atoms results splitting of the Fe-3d orbitals into e$_g$  (d$_{x^2-y^2}$, d$_{z^2}$) and t$_{2g}$ (d$_{xy}$, d$_{yz}$, d$_{zx}$) manifolds. }} Figure \ref{sfig1}(a)-(c) present the contributions of ${J_{ij}}^{e_{g}-e_{g}}$,  ${J_{ij}}^{t_{2g}-t_{2g}}$ and ${J_{ij}}^{t_{2g}-e_{g}}$ for Fe-3d in 4d/Fe/Ir(111) with 4d = Y, Zr, Nb,  Mo, Ru and Rh respectively.  

\par Figure \ref{sfig2}(a)-(b) show calculated energies, magnetic states with external magnetic field from 0 to 12 T for Nb/Fe/Ir(111) and Rh/Fe/Ir(111) from Monte Carlo simulations with temperature annealing.  Nb/Fe/Ir(111) has a spin spiral phase with period of 9 Fe atoms (see inset figures of  \ref{sfig2}(a)) as a magnetic ground state which is robust against external magnetic field.  Whereas the combination of spin spiral and isolated skyrmion appears as a meta stable phase in Rh-Fe/Ir(111) which flips into skyrmion lattice phase with external magnetic field of $\approx$ 18 T.  The wave length of SS in Nb/Fe/Ir(111) (2.28 Å) are much larger compared to Rh/Fe/Ir(111) (1.61 Å).  $\frac{|{D_{ij}}|}{\bar{J_{ij}}}$ decreases much in  Rh/Fe/Ir(111) compared to others 4d/Fe/Ir(111) which causes the magnetic phase transition of ground state from SS to SkL in Rh/Fe/Ir(111) with external magnetic field.  The energy of the SkL phase reduces with external magnetic field and got lower energy value compared to its SS phase above $\approx$ 18 T.  This is the reason for magnetic phase transition from SS to SkL in Rh/Fe/Ir(111) compared to others 4d/Fe/Ir(111) as well as Nb/Fe/Ir(111).

\par
\begin{figure*} [ht] 
\centering
\includegraphics[width=0.99\textwidth,angle=0]{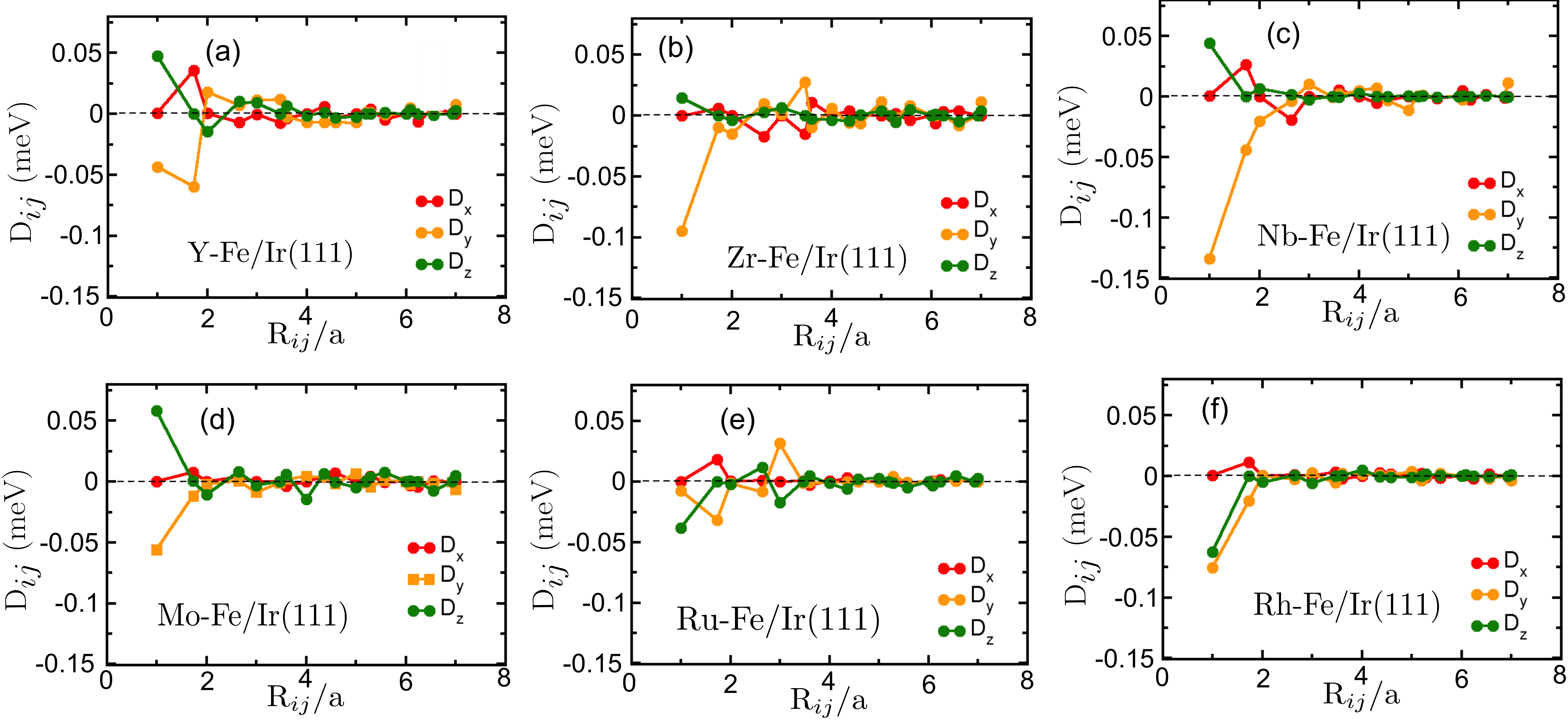} 
\caption{{\bs {The components of the anti-symmetric Dzyaloshinskii-Moriya interactions for 4d/Fe/Ir(111) transition metal multilayers.}}}
\label{sfig3} 
\end{figure*}

 {\bs {Figure \ref{sfig3} (a)-(f) represent  x, y,  z component of the anti-symmetric Dzyaloshinskii-Moriya (DM) interactions to interpret the origin of the strength of the DM vectors for 4d/Fe/Ir(111) transition metal multilayers.  Here we choose only one first nearest neighbour (NN) out of six first NN,  one second NN out of six second NN and so on relating to the Fig.\ref{fig2} for 4d/Fe/Ir(111) including Rh/Fe/Ir(111).  For the first and second NN,  D$_x$ appear as zero and positive respectively,  {\bl {whereas both are positive}} for D$_y$ components of DM interactions.  The in-plane components of DM interactions follows similar trends for all 4d/Fe/Ir(111) transition metal multilayers whereas the out of plane components (D$_z$) change its sign from positive to negative for Ru-Fe/Ir(111) and Rh/Fe/Ir(111).   D$_y$ and  D$_z$ are negative for Rh/Fe/Ir(111) which results the negative strength of DM vectors for some NN as shown in Fig.\ref{fig2}. }}

\end{document}